\documentclass[12pt, a4paper]{article}
\usepackage{graphicx}

\usepackage{amsmath,amsfonts,latexsym}
\usepackage{color}
\usepackage[usenames,dvipsnames]{xcolor}
\usepackage{graphicx}

\definecolor{bubblegum}{rgb}{0.99, 0.76, 0.8}
\definecolor{carnationpink}{rgb}{1.0, 0.65, 0.79}
\definecolor{cherryblossompink}{rgb}{1.0, 0.72, 0.77}
\definecolor{classicrose}{rgb}{0.98, 0.8, 0.91}
\definecolor{cottoncandy}{rgb}{1.0, 0.74, 0.85}
\definecolor{electriclavender}{rgb}{0.96, 0.73, 1.0}
\definecolor{flamingopink}{rgb}{0.99, 0.56, 0.67}
\definecolor{hotpink}{rgb}{1.0, 0.41, 0.71}
\definecolor{ticklemepink}{rgb}{0.99, 0.54, 0.67}

\definecolor{KMpink}{RGB}{255, 153, 255}
\definecolor{KMgreen}{RGB}{0, 127, 0}

\begin{document}
\bibliographystyle{unsrt}
\vspace{-2.5cm}

\title {Weakly nonlinear extension of d'Alembert's formula}
\author {K.R. Khusnutdinova,
K.R. Moore\thanks{Corresponding author. Tel: +44 (0)1509 228203. Fax: +44 (0)1509 223969.} }
\date{}
\maketitle
\vspace{-5ex}
\begin{center}
Department of Mathematical Sciences,\\
Loughborough University,
Loughborough LE11 3TU, UK\\[2ex]
K.Khusnutdinova@lboro.ac.uk\\
K.R.Moore@lboro.ac.uk

\end{center}

\abstract{We consider a weakly nonlinear solution of the Cauchy problem for the regularised Boussinesq equation, which constitutes an extension of the classical d'Alembert's formula for the linear wave equation. 
The solution is given by a simple and explicit formula, expressed in terms of two special functions solving the initial-value problems for  two Korteweg-de Vries equations. We test the formula by considering several examples with `exactly solvable initial conditions' and their perturbations. Explicit analytical solutions are compared with the results of direct numerical simulations.}
\bigskip


\bigskip

\section{Introduction}

The Boussinesq equation and its generalisations, since the original derivation in the context of fluids \cite{BOUS} and reappearance in connection with the famous Fermi-Pasta-Ulam problem \cite{FPU,ZK},
have recently emerged in a vast variety of problems describing nonlinear waves in solids (see, for example,  \cite{Christov,Maugin,Samsonov,Porubov,JE,Fission,KSZ}). 
At the same time, considerable progress has been made in understanding the validity of regularised models and proofs of existence and local well-posedness of the initial-value problem (IVP) in the context of water waves (see \cite{BBM,BONA1,BONA2,BCL,Lannes} and references therein), while the global well-posedness is known to be a complicated issue \cite{KL}. Some progress also has been made in the study of boundary-value problems \cite{FP}. Most relevant to our present paper are the results establishing the validity of two Korteweg-de Vries equations (KdV) as a leading order approximation to solutions of Boussinesq-type equations, as well as some results for the higher order corrections \cite{Gear,Craig,KN,K,J,S,BYC,SW,WW}.

Recently, we constructed a weakly nonlinear solution of the initial-value problem for a system of coupled Boussinesq equations on the infinite line  \cite{KM}, assuming that initial data generates sufficiently rapidly decaying right- and left-propagating waves, in terms of solutions of various Ostrovsky-type equations (see \cite{Ostrovsky} for the original Ostrovsky equation). When coupling parameters are equal to zero, our results yield a simple formula for a weakly nonlinear solution  of an IVP for the regularised Boussinesq equation 
\begin{eqnarray}
f_{tt} - f_{xx} = \epsilon \left [\frac 12 (f^2)_{xx} + f_{ttxx}\right ] + O(\epsilon^2),
\label{eqn:bous}
\end{eqnarray}
also known as the improved Boussinesq equation. The advantage of using model (\ref{eqn:bous}) is that it obviates the short-wave instability (see \cite{BBM}), which is inherent in other versions of the Boussinesq equation, although this version is not integrable \cite{ZAK,AS}.  
%


The paper is organised as follows. In Section  \ref{sec:WNS}  we firstly generalise our formula obtained in \cite{KM} by considering the case when initial conditions are split into $O(1)$ and $O(\epsilon)$ terms, allowing one to construct explicit analytical solutions for a wider class of initial conditions. We use asymptotic multiple-scale expansions, similar to the type used in a study of oblique interaction of solitary waves \cite{Miles1}, and an averaging, used for example in \cite{BYC,KM}, to derive asymptotically exact models which describe the leading order terms in the expansions. The leading order terms satisfy the IVP for two KdV equations \cite{KDV}, integrable by the Inverse Scattering Transform (IST) \cite{GGKM} (see also \cite{AS,DJ}). We use two arbitrary functions present in the higher order terms to improve the accuracy of the approximate solution. The derived formula constitutes a weakly nonlinear extension of the classical d'Alembert's formula for the linear wave equation, and has a similar structure. To the best of our knowledge such a formula was not suggested in previous studies.  The simple and explicit form of this solution allows us to construct a large class of approximate solutions corresponding to `exactly solvable initial conditions' (see Section \ref{sec:examples}).
The weakly nonlinear approach used in this paper can be applied to any form of the Boussinesq equation and coupled systems of Boussinesq equations (see \cite{KM})  and, in particular, it offers an alternative to implementing the IST to the integrable version of the equation in physically relevant cases. 

%
%
In Section \ref{sec:numerics} we introduce a finite difference scheme to solve a scaled version of (\ref{eqn:bous}), in the form
\begin{eqnarray}
\tilde{f}_{\tilde{t}\tilde{t}}- \tilde{f}_{\tilde{x}\tilde{x}} = \frac 12 (\tilde{f}^2)_{\tilde{x}\tilde{x}} + \tilde{f}_{\tilde{t}\tilde{t}\tilde{x}\tilde{x}} ,
\label{eqn:bous_num}
\end{eqnarray}
where $
\tilde{f} =  \epsilon f , \  t = \sqrt{\epsilon}\  \tilde{t}, \ x = \sqrt{\epsilon}\  \tilde{x}
$. 
Equation (\ref{eqn:bous_num}) has particular solitary wave solutions of the form
\begin{eqnarray}
\tilde{f} = A\  {\rm sech}^2 \left(\frac{\tilde{x} - v \tilde{t}}{\Lambda} \right), \quad \mbox{where} \quad A = 3 (v^2 - 1), \quad \Lambda = \frac{2 v}{\sqrt{v^2 - 1}}. 
\label{eqn:soliton}
\end{eqnarray}
%
We analyse and discuss the schemes stability and directly measure its accuracy by comparing the analytical solution (\ref{eqn:soliton}) (denoted $f_{sol}$) to corresponding numerical simulations ($f_{num}$), at times within the derived asymptotic models validity (i.e. from $\tilde t = 0$ to some point between $\tilde{t} \approx t_a=\frac{1}{\epsilon \sqrt{\epsilon}}$ and $\tilde{t} \approx t_b=\frac{1}{\epsilon ^2\sqrt{\epsilon}}$). 

In Section \ref{sec:examples} we explicitly derive the weakly nonlinear solution for exactly solvable initial conditions of the IVP in the form of right-propagating and both right- and left-propagating N-soliton solutions of the KdV equation. Note that these initial conditions do not correspond to the exact soliton solutions of the Boussinesq equation.

In Section \ref{sec:partic_examples} we consider particular cases of initial conditions, namely one- and two-soliton solutions of the KdV equation, and analyse the absolute error in comparison with relevant numerical simulations. We define the maximum error over $\tilde{x}$ of the solution at $\tilde{t}=\tau$ as
\begin{eqnarray*}
e^l_{\tau} = { \max_{-L \leq \tilde{x} \leq L}}  | f_{num}(\tilde{x},\tau) - \tilde{f}_l(\tilde{x},\tau)|, \quad {\rm for } \ \ l=1,2,
\label{eqn:error}
\end{eqnarray*}
where we restrict $\tilde{x}$ to the finite domain $2L$ and $\tilde{f}_l(\tilde{x},\tilde{t})$ denotes the weakly nonlinear solution up to and including $O(\epsilon^{l})$ terms. 

In Section \ref{sec:1way_nsol_pert} we consider perturbations to exactly solvable initial conditions of the IVP, in particular for the case of right-propagating N-soliton solutions, and again analyse the error for a specific example. We finish with concluding remarks in Section 7.
%
%
%
%
%
\newpage

\section{Weakly nonlinear solution} \label{sec:WNS}

We consider the following IVP for  a single regularised Boussinesq equation on the infinite line
\begin{eqnarray}
&&f_{tt} - f_{xx} = \epsilon \left [\frac 12 (f^2)_{xx} + f_{ttxx}\right ] + O(\epsilon^2), \nonumber \\
&& f|_{t=0} = F^0(x)+ \epsilon F^1(x) + O(\epsilon^2),  \ \ \ \ f_t|_{t=0} = V^0(x) + \epsilon V^1(x) + O(\epsilon^2),
\label{eqn:IVP}
\end{eqnarray}
(or any other asymptotically equivalent form of this equation) for the case when the initial conditions generate sufficiently rapidly decaying right- and left-propagating waves. Therefore to leading order the initial ($t=O(1)$) evolution of the Cauchy data is described by the classical d'Alembert's solution
\begin{eqnarray*}
f_0 (x,t) = f_0^- (x-t) + f_0^+ (x+t)         \qquad {\rm for}     \qquad       f_0^{\pm} (x \pm t) = \frac 12 \left ( F^0(x \pm t) \pm \int_{-\infty}^{x \pm t} V^0(x) dx \right ). 
\label{eqn:f0}
\end{eqnarray*}
In general $f_{0}^{\pm}$ are some step like functions but in what follows we shall restrict our considerations to the case when these functions are sufficiently rapidly decaying at infinity (in particular, $\int_{-\infty}^{\infty} V(x) dx = 0$). 

To describe the subsequent $(t=O(\epsilon^{-1}))$ evolution of the Cauchy data we introduce the slow time variable $T=\epsilon t$ and seek the following weakly nonlinear solution in the form of asymptotic multiple scale expansions
\begin{eqnarray}
 f = f^-(\xi, T) + f^+(\eta, T) + \epsilon f^1(\xi, \eta, T) + O(\epsilon^2),
\label{eqn:WNS}
\end{eqnarray}
where $\xi = x-t$ and $\eta = x+t$. Substituting (\ref{eqn:WNS}) into the Boussinesq equation (\ref{eqn:IVP}) we find at leading order the equation is satisfied, whilst at $O(\epsilon)$ we obtain
\begin{eqnarray}
-4 f^1_{\xi \eta} &=& (2 f^-_T + f^- f^-_{\xi} + f^-_{\xi \xi \xi})_\xi 
 +  (-2 f^+_T  + f^+ f^+_{\eta} + f^+_{\eta \eta \eta} )_\eta   +   2 f^-_\xi f^+_\eta  + f^+ f^-_{\xi \xi} + f^- f^+_{\eta \eta} .
 \label{eqn:sub_f1}
 \end{eqnarray}
The subsequent derivation can be performed either by integrating (\ref{eqn:sub_f1}) and requiring that $f^1$ is nonsecular (see, for example, \cite{Gear}) or using averaging arguments (see, for example, \cite{BYC,KM}). In what follows we use the latter of the two. Namely, we average (\ref{eqn:sub_f1}) with respect to the fast time variable $t$ at constant $\xi$ or $\eta$; the left hand side of (\ref{eqn:sub_f1}) at constant $\xi$ is averaged as follows
\begin{eqnarray*}
\lim_{\tau \to \infty} \frac{1}{\tau} \int_0^\tau f^1_{\xi \eta} dt = \lim_{\tau \to \infty} \frac{1}{2\tau} \int_{\xi}^{\xi + 2\tau} f^1_{\xi \eta} d \eta =  \lim_{\tau \to \infty} \frac{1}{2\tau} \left [f^1_\xi \right ]_{\eta=\xi}^{\eta=\xi+2\tau} =0,
\end{eqnarray*}
where we assume $f^1$ and its derivatives remain bounded (required to have a nonsecular expansion (\ref{eqn:WNS})), and similarly we get zero when averaging the same term at constant  $\eta$. Averaging entirely over (\ref{eqn:sub_f1}) at constant $\xi$ and assuming $f^{\pm}$ and their derivatives remain bounded and are sufficiently rapidly decaying at infinity for any fixed $T$ (consistent with relevant numerical experiments), we obtain
\begin{eqnarray*}
0 & = &\left (2 f^-_T +  f^- f^-_{\xi} + f^-_{\xi \xi \xi}\right )_\xi  
\nonumber \\
 &&  +   \lim_{\tau \to \infty} \frac{1}{2\tau}   \left\{    \left[ -2 f^+_T + f^+ f^+_{\eta} + f^+_{\eta \eta \eta}\right]^{\eta=\xi + 2\tau}_{\eta=\xi}     
+  \int^{\xi + 2\tau}_{\xi}  ( 2 f^-_\xi f^+_\eta  + f^+ f^-_{\xi \xi} + f^- f^+_{\eta \eta} ) d\eta    \right\} \nonumber \\
& = & \left (2 f^-_T +  f^- f^-_{\xi} + f^-_{\xi \xi \xi}\right )_\xi  .
 \label{O1a}
\end{eqnarray*}
Averaging entirely over (\ref{eqn:sub_f1}) at constant $\eta$ we derive a similar equation for  the function $f^+$.
 Integrating each of these equations with respect to their respective characteristic variables, and taking into account the behaviour  of $f^{\pm}$ at infinity, yields the following two KdV equations
\begin{eqnarray}
f^-_T + \frac 12 f^- f^-_\xi + \frac 12 f^-_{\xi \xi \xi} = 0, \qquad f^+_T - \frac 12 f^+ f^+_\eta - \frac 12 f^+_{\eta \eta \eta} = 0.
\label{eqn:orig_KDV}
\end{eqnarray}
The higher order correction $f^1$ is then obtained by substituting (\ref{eqn:orig_KDV}) into (\ref{eqn:sub_f1}) to yield
\begin{eqnarray}
f^1 = - \frac 14 \left (2 f^- f^+  + f^-_{\xi} \int f^+ d\eta + f^+_{\eta} \int f^- d\xi  \right )+ \phi(\xi, T) + \psi(\eta, T).
\label{eqn:WNS_higher}
\end{eqnarray}
Finally substituting the weakly nonlinear solution (\ref{eqn:WNS}) into the initial conditions of the IVP (\ref{eqn:IVP}), we derive at leading order initial conditions with respect to $T$, for the leading order terms $f^{\pm}$
\begin{eqnarray*}
f^{\pm}|_{T=0} = f_0^{\pm}   = \frac 12 \left ( F^0(x \pm t) \pm \int_{-\infty}^{x \pm t} V^0(x) dx \right ),
\label{eqn:orig_KDV_ICs}
\end{eqnarray*}
whilst at $O(\epsilon)$ we obtain, within the accuracy of the problem formulation, the following  d'Alembert's-like formulae for the functions $\phi(\xi,T)$ and $\psi(\eta,T)$:
\begin{eqnarray}
\phi(\xi, T) = \frac 12 \left [R_{1} (\xi, T) + \int_{-\infty}^\xi R_{2} (x, T) dx\right ],   \quad   \ \
\psi(\eta, T) = \frac 12 \left [R_{1} (\eta, T) - \int_{-\infty}^\eta R_{2} (x, T) dx\right ], 
\label{eqn:phi_psi}
\end{eqnarray}
where
\begin{eqnarray}
&&R_{1} (x, T) = \frac 14 \left [2 f^{-} f^{+} + f^-_{\xi} \int f^+ d\eta + f^+_{\eta} \int f^- d\xi  \right ]_{t=0} + F^1(x),  \nonumber  \\
&&R_{2} (x, T) = \left [f^{-}_T + f^{+}_T + \frac 14 \left (f^+ f^{-}_{\xi} - f^{-} f^{+}_{\eta} +  f^-_{\xi \xi} \int f^+ d\eta - f^+_{\eta \eta} \int f^- d\xi  \right  )\right ]_{t=0}  - V^1(x).
\label{eqn:P_Q}
\end{eqnarray}
The dependence of the functions $\phi$ and $\psi$ on $T$ is inherited from their dependence on the leading order functions $f^-$ and $f^+$, however this may be neglected, at least for sufficiently small values of time. To construct a more accurate solution, valid for greater values of time, one needs to know higher-order terms in the problem formulation (\ref{eqn:IVP}).

%
%
%
%
%
%
\section{Numerical scheme} \label{sec:numerics}
We next discuss the numerical scheme used to solve the Boussinesq equation (\ref{eqn:bous_num}). We implement a finite difference scheme derived in \cite{SOR} for a regularised Boussinesq equation or equivalently the scheme used in \cite{KM} for the reduction $g=\delta=\gamma=0$ and $c=\alpha=\beta=1$.

We let $\tilde{x}\in [-L,L]$, for finite $L$, and discretise the $(\tilde{x},\tilde{t})$ domain into a grid with spacings $\Delta \tilde{x}=h$ and $\Delta \tilde{t}=\kappa$. The solution $\tilde{f}(\tilde{x},\tilde{t})$ of (\ref{eqn:bous_num}) is approximated by the solution $f(ih,j\kappa)$ (for $i=0,1,...,N$ and $j=0,1,...$) of the difference scheme, denoted $f_{i,j}$.

Substituting central difference approximations into equation (\ref{eqn:bous_num}) we derive the following difference scheme
\begin{eqnarray}
 - f_{i-1,j+1} + (2 +h^2)f_{i,j+1}-f_{i+1,j+1} &=& (\kappa^2-2)[f_{i-1,j}-2f_{i,j}+f_{i+1,j}] + 2h^2f_{i,j} 
 \nonumber \\
&& + \frac{\kappa^2}{2}\left[(f_{i-1,j})^2-2(f_{i,j})^2+(f_{i+1,j})^2\right]
\nonumber
\\
&& + f_{i-1,j-1} - (2 +h^2)f_{i,j-1}+ f_{i+1,j-1}.
\label{eqn:scheme}
\end{eqnarray}
We choose the boundary conditions such that they are sufficiently far away from the propagating waves. Periodic boundary conditions were also considered but, for the added complexity and computational effort involved in solving the difference scheme, the difference in the error of the solution was negligible. Thus we set
 \begin{eqnarray*}
f_{0,j}=f_{N,j}=0, \quad \forall j.
\label{eqn:BCs}
\end{eqnarray*}
Initial conditions for simulations of the scheme are chosen to coincide with initial conditions in the IVP (\ref{eqn:IVP}). However for the purpose of analysis of error in this section we choose the initial conditions in the form of the particular Boussinesq soliton solution (\ref{eqn:soliton}) and thus
\begin{eqnarray*}
f_{i,0} = A\  \text{sech} ^2\left(\frac{\tilde{x}}{\Lambda}\right), \qquad 
f_{i,1} = A\  \text{sech} ^2\left(\frac{\tilde{x}-v\kappa}{\Lambda}\right),  \qquad \forall i,
\label{eqn:ICs}
\end{eqnarray*}
where $A=3(v^2-1), \ \Lambda=2v(v^2-1)^{-\frac{1}{2}}$ and we choose $v\approx\sqrt{\frac{\epsilon}{3}+1}$ to ensure the amplitude is $O(\epsilon)$ (required to make comparisons with the weakly nonlinear solution in the next sections). The nine point implicit difference scheme (\ref{eqn:scheme}), with tri-diagonal matrices of constant coefficients, is solved using a Thomas Algorithm (e.g., \cite{thomas}). 

To examine the schemes stability we firstly linearise by setting $f_{i,j}=f_0+\hat{f}_{i,j}$, where $f_0$ is a constant such that $f_0>f_{i,j}$ $\forall i,j$. Then using a Von-Neumann stability analysis we substitute $\hat{f}_{i,j}=G^je^{\text{i}\theta ih}$ (where $\text{i}^2 = -1$) into the linearised version of (\ref{eqn:scheme}) and derive the condition 
\begin{eqnarray*}
G^2-2\mu G+1=0\ \ \ {\rm{where}} \ \ \  \mu=1-\frac{2\kappa^2 (1 +f_0) {\rm{sin}}^2 \frac{\theta h}{2}}{h^2 +4  {\rm{sin}}^2 \frac{\theta h}{2}}.
\label{eqn: G}
\end{eqnarray*}
For stability we require $|G|\leq1$ $\forall\theta$ and arbitrary $\kappa,h$, which is true provided $|\mu|<1$ and thus implies 
\begin{eqnarray}
\kappa<\kappa_c=\sqrt{\frac{h^2+4}{1+f_0}}. 
\label{eqn:criteria}
\end{eqnarray}
Hence the roots of the quadratic in $G$ have modulus one and the linearised form of  the difference scheme (\ref{eqn:scheme}) is stable provided $\kappa<\kappa_c$. (In practice, we used a stricter condition $\kappa < \frac 12 \kappa_c$, to accommodate for the effects of nonlinearity). It can be shown that the principal truncation error of the linearised scheme is $O(h^2\kappa^4 + h^4\kappa^2)$.

For a given $h$ the restriction (\ref{eqn:criteria}) can be used to determine valid choices of $\kappa$ to ensure stability. Table 1 displays the maximum absolute error across $\tilde{x}$ for different choices of $\kappa$ for fixed $ \epsilon$ (and hence fixed $v$) for two different space step sizes $h$, and at two different times $t_a$ and $t_b$ (within the asymptotic region of validity of the KdV models (\ref{eqn:orig_KDV})). The optimal discretisations $h=\kappa=0.1$ are chosen for simulations where the maximum absolute error within the time interval considered ranges from $O(10^{-6})$ to $O(10^{-7})$. This magnitude of error of the numerical solution is deemed suitable from previous numerical studies on Boussinesq-type equations (for example, see \cite{P11}, \cite{P3}, \cite{P4}, \cite{P9}, \cite{P6}).

\begin{table*} [htbp]
\centering
\begin{tabular}{c  c  c  c  c }  \hline \vspace{-4.5mm} \\ 
&\multicolumn{2}{c}{{\rm Max\ error\ at\ $\tilde{t}=32$}} &  \multicolumn{2}{c}{\rm Max\ error\ at\ $\tilde{t}=300$}  \\ \hline 
$\kappa$ & h=0.1  & h=0.01 & h=0.1  & h=0.01 \\ \hline
1& \  \  4.4693 ${\rm x \ 10^{-4}}$  \ \   &  \  \  4.5089 ${\rm x \ 10^{-4}}$   \  \  &  \ \ \ \ 3.3 ${\rm x \ 10^{-3}}$  &  \ \ \ \ 3.3 ${\rm x \ 10^{-3}}$   \\ 
0.5&  8.8851 $\rm{x} \ 10^{-5}$ & 1.1299 ${\rm x \ 10^{-4}}$ &  \  \  7.7082 ${\rm x \ 10^{-4}}$   \ \ & \ \ 8.0012 ${\rm x \ 10^{-4}}$ \ \ \\
0.2& 1.443 $\rm{x} \ 10^{-5}$ & 1.8105 ${\rm x \ 10^{-5}}$ &  9.7890 ${\rm x \ 10^{-5}}$  & 1.2671 ${\rm x \ 10^{-4}}$ \\
0.1& {\bf 5.2846} ${\rm\bf{{x} \ 10^{-7}}}$ & 4.5133 ${\rm x \ 10^{-6}}$ &  {\bf 2.6891} ${\rm {\bf x \ 10^{-6}}}$  & 3.1518 ${\rm x \ 10^{-5}}$ \\
0.05& 2.8857 $\rm{x} \ 10^{-6}$ & 1.1439 ${\rm x \ 10^{-6}}$ &  2.1081 ${\rm x \ 10^{-5}}$  & 8.0360 ${\rm x \ 10^{-6}}$ \\
0.025& 3.7396 $\rm{x} \ 10^{-6}$ & 4.3677 ${\rm x \ 10^{-7}}$ &  2.7026 ${\rm x \ 10^{-5}}$ & 3.2910 ${\rm x \ 10^{-6}}$ \\
0.0125& 3.9559 $\rm{x} \ 10^{-6}$ & 9.0239 ${\rm x \ 10^{-7}}$ & 2.8532 ${\rm x \ 10^{-5}}$  & 1.1751 ${\rm x \ 10^{-5}}$  \\
0.00625& 4.0198 $\rm{x} \ 10^{-6}$ & 3.6362 ${\rm x \ 10^{-6}}$ & 2.8988 ${\rm x \ 10^{-5}}$  & 4.6968 ${\rm x \ 10^{-5}}$  \\ \hline
\end{tabular}
\caption{\small Maximum absolute error of the numerical solution compared to the exact solution (\ref{eqn:soliton}) for various time and space discretisations $h$ and $\kappa$, with $\epsilon=0.1$.}
\end{table*}

\begin{figure}[htbp]
\begin{center}$
\begin{array}{ccc}
\quad \tilde{t}=32 & \quad \tilde{t}=300  \\ \\
\includegraphics[width=2.4in]{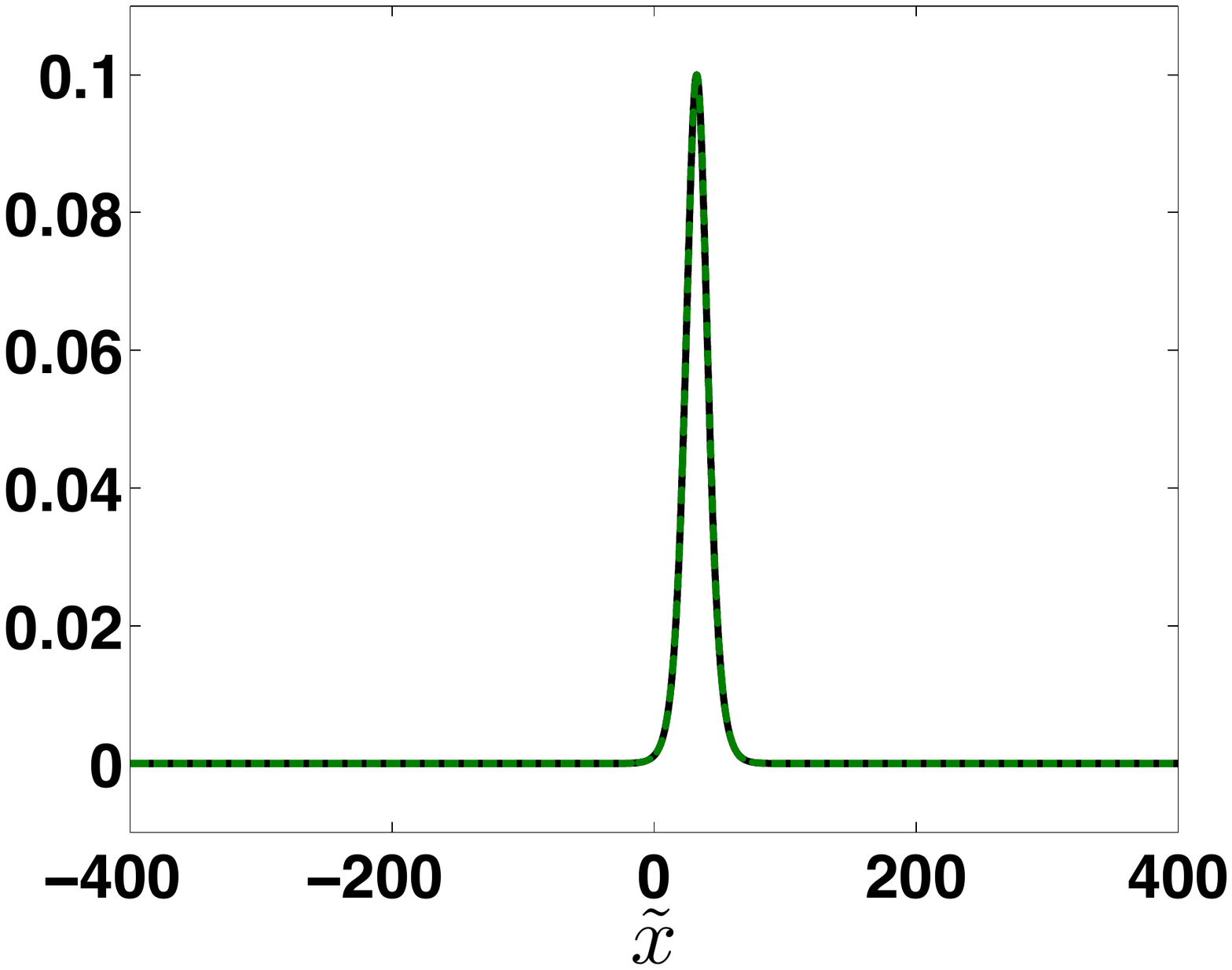} &
\includegraphics[width=2.4in]{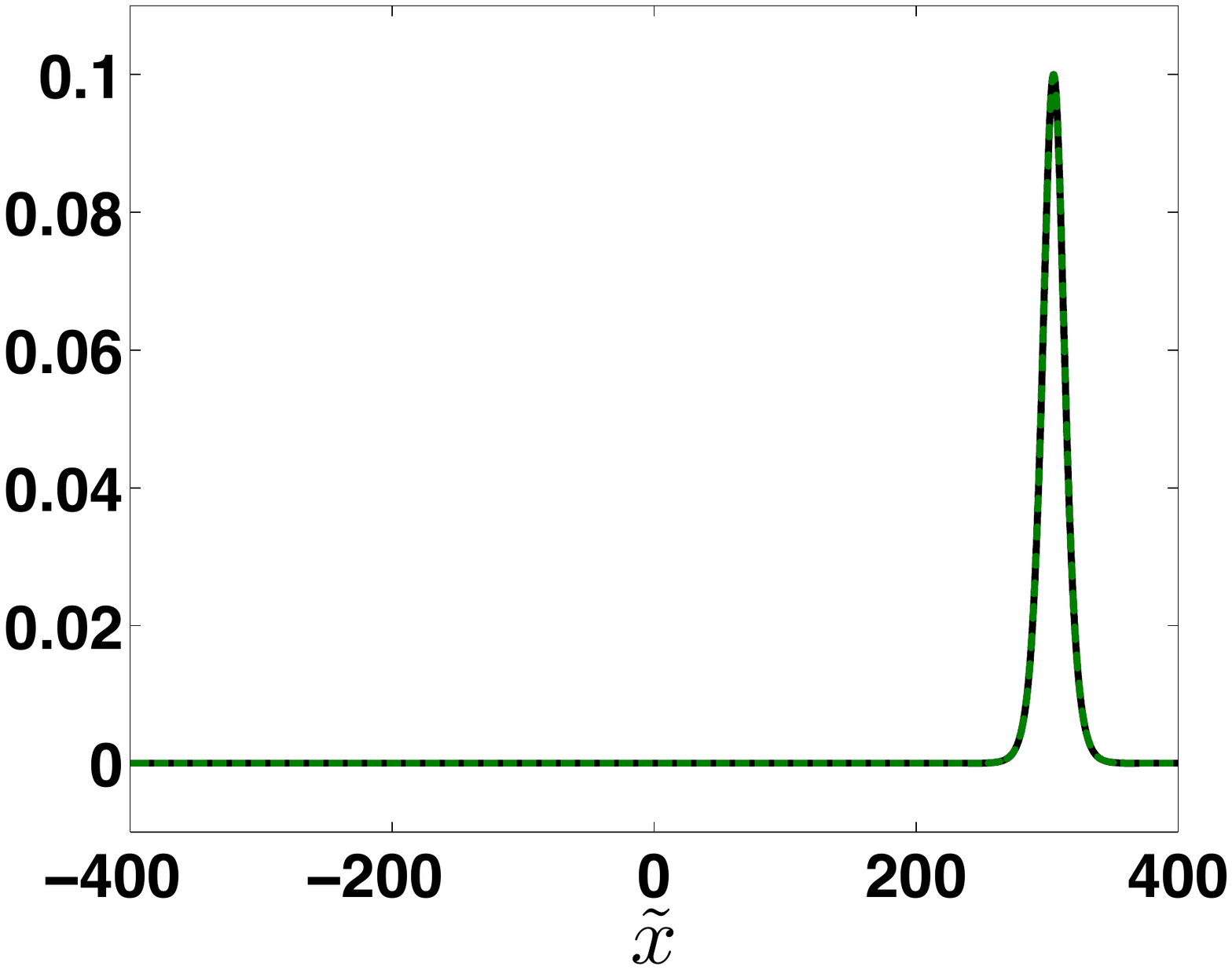}  \\
\quad  \mbox{\footnotesize \bf (a) $f_{{\rm num}}$({\color{black} \bf ---}) \  $f_{{\rm sol}}$({\color{KMgreen} \bf - -}) } &
\quad  \mbox{\footnotesize \bf (b) $f_{{\rm num}}$({\color{black} \bf ---}) \  $f_{{\rm sol}}$({\color{KMgreen} \bf - -}) }  \\
\includegraphics[width=2.4in]{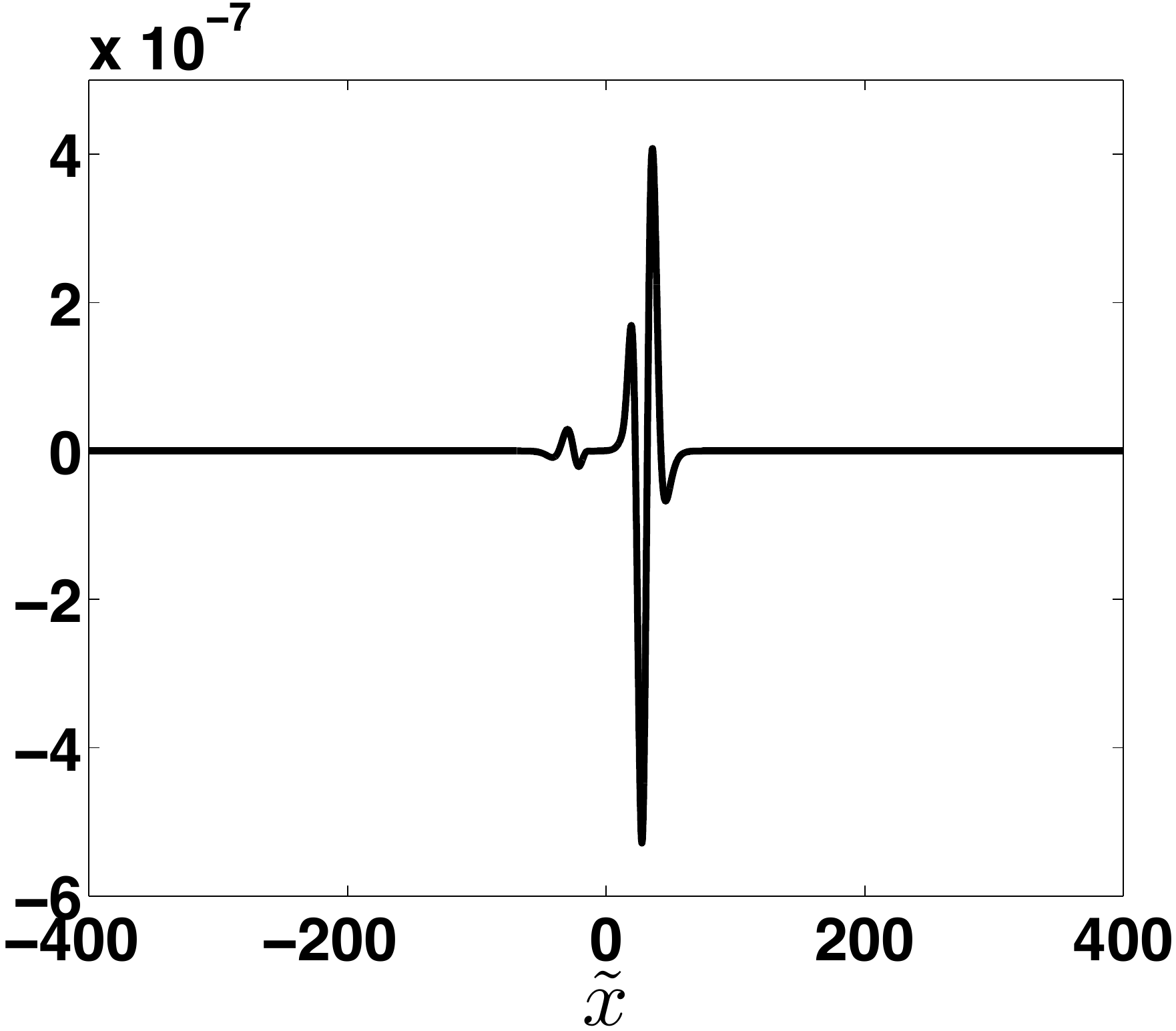}  &
\includegraphics[width=2.4in]{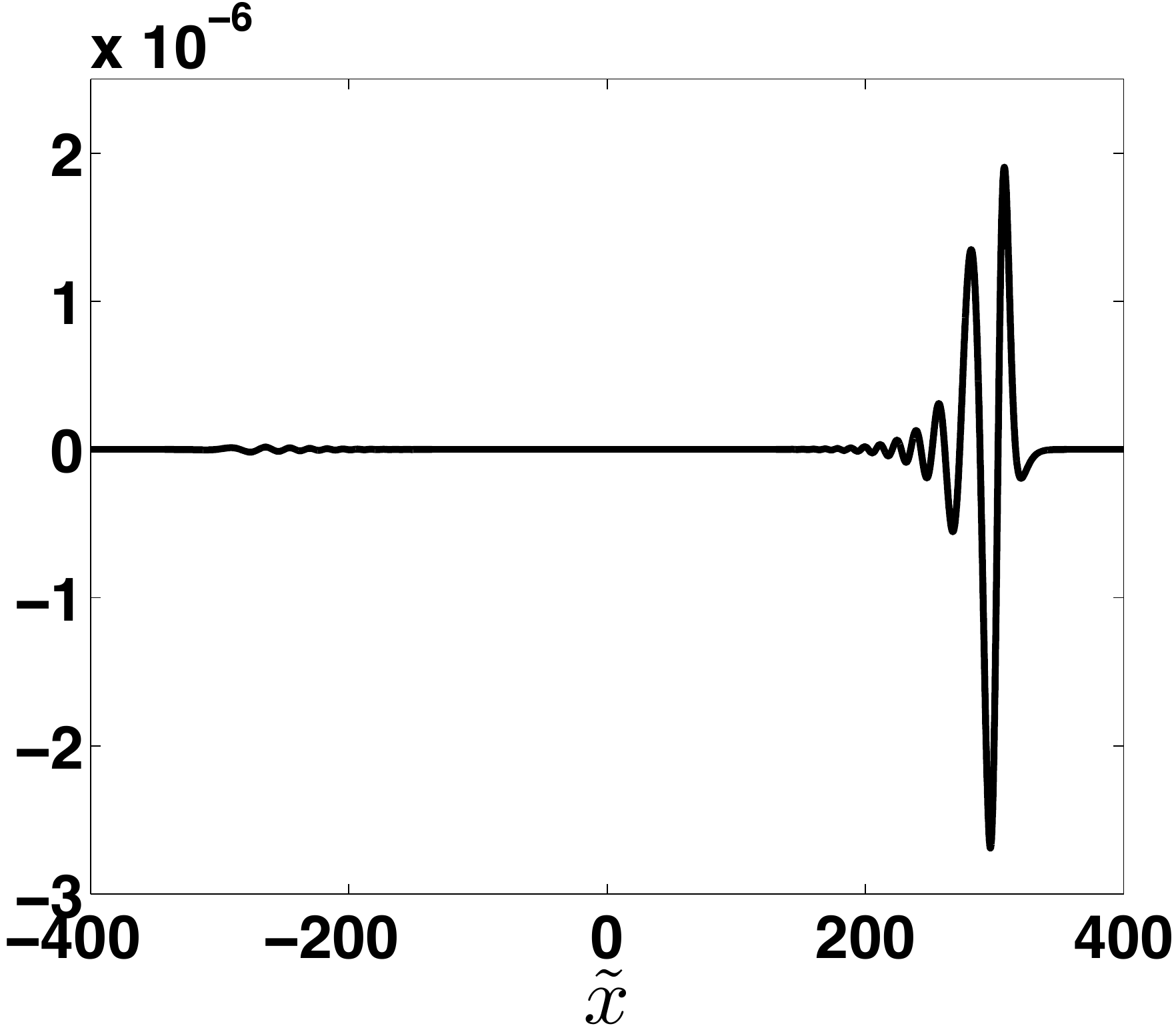} \\
\mbox{ \footnotesize \bf (c) \footnotesize  ($f_{{\rm num}}$  -  $f_{{\rm sol}}$)({\color{black} \bf ---}) }  &
\mbox{ \footnotesize \bf (d) \footnotesize  ($f_{{\rm num}}$  -  $f_{{\rm sol}}$)({\color{black} \bf ---})  }
\end{array}$
\end{center}
\caption{\small Evolution of numerical solution and exact solution (\ref{eqn:soliton}) at (a) $\tilde{t}=32\ (\approx t_a) \ \& \ $(b) $\tilde{t}=300\ (\approx t_b)$, and the absolute errors (c) $\&$ (d) at the respective times. Parameter values are $h=\kappa=0.1$ and $\epsilon=0.1$.}
\label{figure:num_anal_error}
\end{figure}

Figure \ref{figure:num_anal_error} shows the evolution of the numerical solution of (\ref{eqn:scheme}) $(f_{num})$ compared with the particular analytical solution (\ref{eqn:soliton}) $(f_{sol})$, along with the respective error plots at each time. It's clear that for this range of time, the numerical solution is in excellent agreement with the analytical solution.

\pagebreak

Finally Table (\ref{table:numerics_error_eps}) depicts the maximum absolute error over $\tilde{x}$ for varying $\epsilon$ at the corresponding $\tilde{t}\approx t_a$. This measure of accuracy for the scheme is utilised when analysing the error of the weakly nonlinear solution for particular examples in Sections \ref{sec:partic_examples} and \ref {sec:1way_nsol_pert}.
 
\begin{table} [htbp]
\centering
\begin{tabular}{ c  c  c } \hline  \vspace{-4.5mm} \\ 
$\epsilon$ & $\tilde{t}(\approx t_a)$  & {\rm Max\ error\ of\ $|f_{num}-f_{sol}|$}  \\ \hline
0.1& 32  &   5.2846 ${\rm x \ 10^{-7}}$    \\ 
0.075& 49 &   3.6296 ${\rm x \ 10^{-7}}$    \\
0.05& 90 &   1.3789 ${\rm x \ 10^{-7}}$      \\
0.025& 253  &  3.4930 ${\rm x \ 10^{-8}}$     \\
0.0125&  716 &  8.4222  ${\rm x \ 10^{-9}}$      \\
0.00625& 2024 &  2.3201 ${\rm x \ 10^{-9}}$     \\ \hline
\end{tabular}
\caption{\small Maximum absolute error of the numerical solution compared to the exact solution (\ref{eqn:soliton}) at $\tilde{t}\approx t_a$ for $h=\kappa=0.1$ and various $\epsilon$.}
\label{table:numerics_error_eps}
\end{table}

%
%
%
%
%
%
\section{Exactly solvable initial conditions}   \label{sec:examples}

Due to the construction of our weakly nonlinear solution it is favourable to split the initial conditions of the IVP (\ref{eqn:IVP}), if possible, into an $O(1)$ `exactly solvable part' (from the viewpoint of the leading order KdV equations) and an $O(\epsilon)$ perturbation. In this section we consider the case when the perturbations are zero. Firstly we transform the KdV equations (\ref{eqn:orig_KDV}) into the standard form
\begin{eqnarray}
\hat{f}^-_{\hat{T}} - 6  \hat{f}^- \hat{f}^-_{\hat{\xi}} + \hat{f}^-_{\hat{\xi} \hat{\xi} \hat{\xi}} = 0, \quad   \quad
\hat{f}^+_{\hat{T}^+} - 6  \hat{f}^+ \hat{f}^+_{\hat{\eta}} + \hat{f}^+_{\hat{\eta} \hat{\eta} \hat{\eta}} = 0,  \label{eqn:new_KDV} \quad \mbox{where} 
\end{eqnarray}
%
\begin{eqnarray*}
\hat{f}^-=-\frac{1}{6\beta^2} f^-, \qquad
\hat{f}^+=-\frac{1}{6\beta^2} f^+, \qquad
\hat{\xi}=\beta \xi, \qquad \hat{\eta}=\beta \eta, \qquad 
 \hat{T}= \frac{\beta^3}{2} T, \qquad 
\hat{T}^+= - \frac{\beta^3}{2} T,
\label{eqn:scaling}
\end{eqnarray*}
and $\beta$ is a free parameter. We consider the class of N-soliton solutions of equations (\ref{eqn:new_KDV}) (see \cite{GGKM,KayM,Hirota}), which  yields, for our original functions, 
%
\begin{eqnarray*}
f^-_N(\xi,T)  = 12 \frac{\partial^2}{\partial \xi^2} {\rm log [ det} M_N(\xi,-T)],   \qquad   
f^+_N(\eta,T)  = 12 \frac{\partial^2}{\partial \eta^2} {\rm log [det} M_N(\eta,T)], 
\label{eqn:f+-_N}
\end{eqnarray*}
for the  $N\ {\rm x}\ N$ matrix $M_N(x,T)=(m_{ij}(x,T))$ with elements
\begin{eqnarray*}
m_{ij}(x,T)  =\delta_{ij} + \frac{2k_i}{k_i+k_j}e^{(k_ix + \frac{k_i^3T}{2} + \alpha_i)} ,
\end{eqnarray*}
where $\alpha_i$ and $k_i$ are arbitrary parameters for $i,j=1...N$ and $\delta_{ij}$ is the Kronecker delta function. %

\subsection{Right-propagating initial conditions} \label{sec:1way_nsol}
Let us firstly consider the case of initial conditions of the IVP (\ref{eqn:IVP}) in the form of right-propagating N-soliton solutions of the KdV equation, i.e.
\begin{eqnarray}
f|_{t=0}= 12 \frac{\partial^2}{\partial x^2} {\rm log [det} M_N(x,0)], \qquad 
f_t|_{t=0} = -12 \frac{\partial^3}{\partial x^3} {\rm log [ det} M_N(x,0)].
\label{eqn:one_way_N_sol_ICs}
\end{eqnarray}
We then have the leading order solutions in the form
\begin{eqnarray*}
f^+=0, \qquad f^-= f^-_N(\xi,T).
\end{eqnarray*}
The higher order terms (\ref{eqn:WNS_higher}) in the weakly nonlinear solution for this case reduce to
\begin{eqnarray*}
f^1 = \phi(\xi,T) + \psi(\eta,T),
\end{eqnarray*}
where
\begin{eqnarray*}
\phi(\xi,T)=\frac 1 2 \int^{\xi}_{-\infty} R_2(x,T) dx,   \qquad     \psi(\eta,T) = -\frac 1 2 \int^{\eta}_{-\infty} R_2(x,T) dx,
\end{eqnarray*}
and $R_2(x,T)=(f_T^-)|_{t=0}$, since all other terms vanish. Therefore we have
\begin{eqnarray*}
f^1\ =\ - \frac 1 2 \frac{\partial}{\partial T} \int^{\eta}_{\xi} f^-_{N}(x,T) dx \ =\ -6\left[\frac{\partial^2}{\partial x \partial T} {\rm log [ det} M_N(x,-T)]  \right]^{x=\eta}_{x=\xi},
\end{eqnarray*}
and consequently the weakly nonlinear solution of the IVP (\ref{eqn:IVP}) for the KdV N-soliton initial conditions propagating to the right, is as follows
\begin{eqnarray}
f =  12 \frac{\partial^2}{\partial \xi^2} {\rm log [det} M_N(\xi,-T)]
-6\epsilon \left[\frac{\partial^2}{\partial x \partial T} {\rm log [ det} M_N(x,-T)]  \right]^{x=\eta}_{x=\xi}     +  O(\epsilon^2).
\label{eqn:WNS_1way}
\end{eqnarray}

\subsection{Right- and left-propagating initial conditions} \label{sec:2way_Nsol}
For the case of right- and left-propagating KdV N-soliton initial conditions of the IVP (\ref{eqn:IVP}), the initial conditions are given by
\begin{eqnarray}
f|_{t=0}= 24 \frac{\partial^2}{\partial x^2} {\rm log [ det} M_N(x,0)], \qquad 
f_t|_{t=0} = 0.
\label{eqn:nsol_both_ICs}
\end{eqnarray}
The leading order solutions take the following form
\begin{eqnarray*}
f^+=f^+_N(\eta,T), \qquad f^-= f^-_N(\xi,T).
\end{eqnarray*}
Unlike the previous case the higher order terms are more difficult to determine since $f^+ \not= 0$. It is convenient to introduce the notation
\begin{eqnarray*}
f^-   =12 U_{\xi\xi}(\xi,T), \qquad f^+=12 V_{\eta\eta}(\eta,T), \quad \mbox{where}
\end{eqnarray*}
$$U(x,T) = {\rm log [det}M_N(x,-T)] \quad \mbox{ and} \quad  V(x,T)= {\rm log [det}M_N(x,T)],$$
 and thus the higher order terms of the weakly nonlinear solution can be written in the form
\begin{eqnarray*}
f^1=-36\left[2U_{\xi\xi}(\xi,T)V_{\eta\eta}(\eta,T)+U_{\xi\xi\xi}(\xi,T)V_{\eta}(\eta,T)+V_{\eta\eta\eta}(\eta,T)U_{\xi}(\xi,T)\right] + \phi(\xi,T) + \psi(\eta,T).
\end{eqnarray*}
The functions $\phi$ and $\psi$ (\ref{eqn:phi_psi}) are constructed from the functions (\ref{eqn:P_Q}), which for this case are in the form
\begin{eqnarray*}
R_1(x,T)&=&36 \frac{\partial}{\partial x} [U_{xx}(x,T)V_{x}(x,T)+U_{x}(x,T)V_{xx}(x,T)], 
\label{eqn:ex2_P} \\
R_2(x,T)&=&12[U_{xxT}(x,T)+V_{xxT}(x,T)]+36  \frac{\partial}{\partial x} \left[U_{xxx}(x,T)V_{x}(x,T) - U_{x}(x,T)V_{xxx}(x,T)\right]. \quad 
 \label{eqn:nsol_both_Q}
\end{eqnarray*} 
It therefore follows that 
\begin{eqnarray*}
\int R_2(x,T)dx = 12[U_{xT}(x,T)+V_{xT}(x,T)]+36[U_{xxx}(x,T)V_{x}(x,T) - U_{x}(x,T)V_{xxx}(x,T)],
\label{eqn:nsol_both_intQ}
\end{eqnarray*} 
and since we have 
\begin{eqnarray*}
\phi+ \psi &=& \frac 1 2 \left[R_1(\xi,T) + R_1 (\eta,T) -  \int^{\eta}_{\xi} R_2(x,T)dx \right]  \nonumber \\   
&=& 36 \lbrace U_{\xi\xi}(\xi,T)V_{\xi\xi}(\xi,T)  +  U_{\eta\eta}(\eta,T)V_{\eta\eta}(\eta,T) + U_{\xi\xi\xi}(\xi,T)V_{\xi}(\xi,T)     +  V_{\eta\eta\eta}(\eta,T)U_{\eta}(\eta,T)    \nonumber \\
&&
+  \frac 1 6 \left[ U_{\xi T}(\xi,T)+V_{\xi T}(\xi,T)  - U_{\eta T}(\eta,T) - V_{\eta T}(\eta,T)  \right]  \rbrace ,
\label{eqn:nsol_both_phi+psI}
\end{eqnarray*} 
the weakly nonlinear solution of the IVP (\ref{eqn:IVP}) for N-soliton KdV initial conditions propagating both left and right, can be explicitly expressed in the form
\begin{eqnarray}
f &=&  12[U_{\xi\xi}(\xi,T) + V_{\eta\eta}(\eta,T)] + 36\epsilon \left \lbrace \frac{\partial}{\partial x}  \left[ U_{\xi\xi}(\xi,T)\left[V_{x}(x,T)\right]^{x=\xi}_{x=\eta} + V_{\eta\eta}(\eta,T) \left[U_{x}(x,T)\right]^{x=\eta}_{x=\xi} \right]  \right. \nonumber \\
&& \left. + \frac{1}{6} \left[  U_{xT}(x,T) +  V_{xT}(x,T) \right] ^{x=\xi}_{x=\eta} \right \rbrace +  O(\epsilon^2). 
\label{eqn:nsol_both_WNS_uv}
\end{eqnarray}  
\section{Examples: weakly nonlinear solution and direct numerical simulations} \label{sec:partic_examples}

\subsection{Right-propagating 1-soliton initial conditions}  \label{sec:1way_1sol}
We now consider particular examples of the two general classes formulated in the previous section and compare them to numerical simulations.
Firstly for the case of initial conditions of the IVP (\ref{eqn:IVP}) in the form of a right-propagating single KdV soliton solution,  the leading order terms are given by
\begin{eqnarray*}
f^+=0, \qquad f^-\ =\ f^-_1(\xi,T) 
\ =\ 12 \frac{\partial^2}{\partial \xi^2} {\rm log}(1+e^{\theta(\xi,-T)}),
\end{eqnarray*}
where $\theta(x,T)=kx + \frac{k^3}{2}T +\alpha$. From (\ref{eqn:one_way_N_sol_ICs}) the initial conditions of the IVP (\ref{eqn:IVP}) are
\begin{eqnarray}
f|_{t=0}= 12 \frac{\partial^2}{\partial x^2} {\rm log} (1+e^{\theta(x,0)}), \qquad 
f_t|_{t=0} = -12 \frac{\partial^3}{\partial x^3} {\rm log} (1+e^{\theta(x,0)})  ,
\label{eqn:1sol_1way_ICs}
\end{eqnarray}
and from (\ref{eqn:WNS_1way}) the weakly nonlinear solution is 
\begin{eqnarray}
f =  12 \frac{\partial^2}{\partial \xi^2} {\rm log} (1+ e^{\theta(\xi,-T)})
-6\epsilon \left[\frac{\partial^2}{\partial x \partial T} {\rm log}  (1+e^{\theta(x,-T)})  \right]^{x=\eta}_{x=\xi}     +  O(\epsilon^2).
\label{eqn:single_WNS_1way}
\end{eqnarray}
Explicitly evaluating the derivatives in (\ref{eqn:single_WNS_1way}) yields
\begin{eqnarray}
f &=& 3k^2{\rm sech}^2\left[\frac{k}{2}\left(\xi - \frac{k^2}{2}T\right) + \frac{\alpha}{2}   \right]\nonumber \\
&+& \frac{3k^4\epsilon}{4}\left \lbrace-{\rm sech}^2  \left[ \frac{k}{2}\left(\xi - \frac{k^2}{2}T\right)  + \frac{\alpha}{2}  \right]
+{\rm sech}^2\left[\frac{k}{2}\left(\eta - \frac{k^2}{2}T\right) + \frac{\alpha}{2} \right] \right \rbrace + O(\epsilon^2).
\label{eqn:ex1_WNS}
\end{eqnarray}
To consider the error of the weakly nonlinear solution for this example, we transform  (\ref{eqn:ex1_WNS}) into the same form used in the numerics
\begin{eqnarray}
\tilde{f} &=& 3k^2\epsilon \  {\rm sech}^2\left[\frac{k \sqrt{\epsilon}}{2}\left(\tilde{x} - \left(1+\frac{k^2\epsilon}{2}\right)\tilde{t}\right)  + \frac{\alpha}{2}  \right] 
+ \frac{3k^4\epsilon^2}{4}\left\{-{\rm sech}^2\left[\frac{k \sqrt{\epsilon}}{2}\left(\tilde{x} - \left(1+\frac{k^2\epsilon}{2}\right)\tilde{t}\right)   + \frac{\alpha}{2}   \right] \right. \nonumber \\
&&+  \left.   {\rm sech}^2\left[\frac{k \sqrt{\epsilon}}{2}\left(\tilde{x} + \left(1-\frac{k^2\epsilon}{2}\right)\tilde{t}\right)  + \frac{\alpha}{2}    \right]\right\}   
+ O(\epsilon^3).
\label{eqn:ex1_WNS_num}
\end{eqnarray}
From (\ref{eqn:1sol_1way_ICs}) we choose the following initial conditions for numerical simulations 
\begin{eqnarray*}
f_{i,0} =3k^2 \epsilon\  \text{sech} ^2 \left(\frac{k\sqrt{\epsilon} {\tilde{x}} + \alpha}{2}\right), \qquad  
f_{i,1} =  3k^2\epsilon\  \text{sech} ^2  \left(\frac{k\sqrt{\epsilon} ({\tilde{x}}-\kappa) + \alpha}{2}\right),    
\end{eqnarray*}
where we choose some $k$ such that the weakly nonlinear solution is applicable. 

\begin{figure}[htbp]
\begin{center}
\includegraphics[width=9cm]{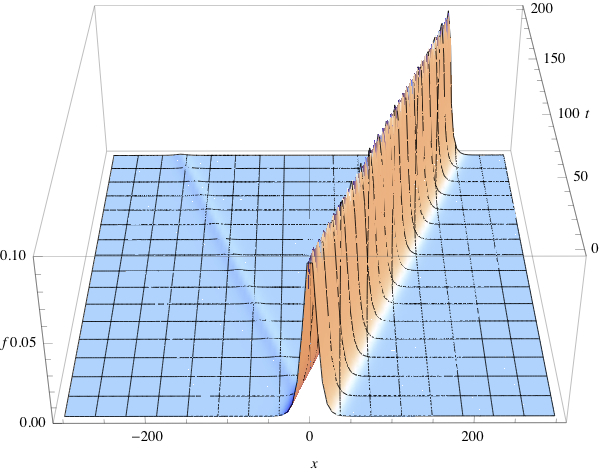}
\caption{\small Evolution of the weakly nonlinear solution for right-propagating 1-soliton initial conditions, with $\epsilon=0.1,\ k=\frac{1}{\sqrt{3}}$ and $\alpha=0$.}
\label{figure:1sol_1way_3D}
\end{center}
\end{figure}

\begin{figure}[htbp]
\begin{center}$
\begin{array}{ccc}
\includegraphics[width=2.45in]{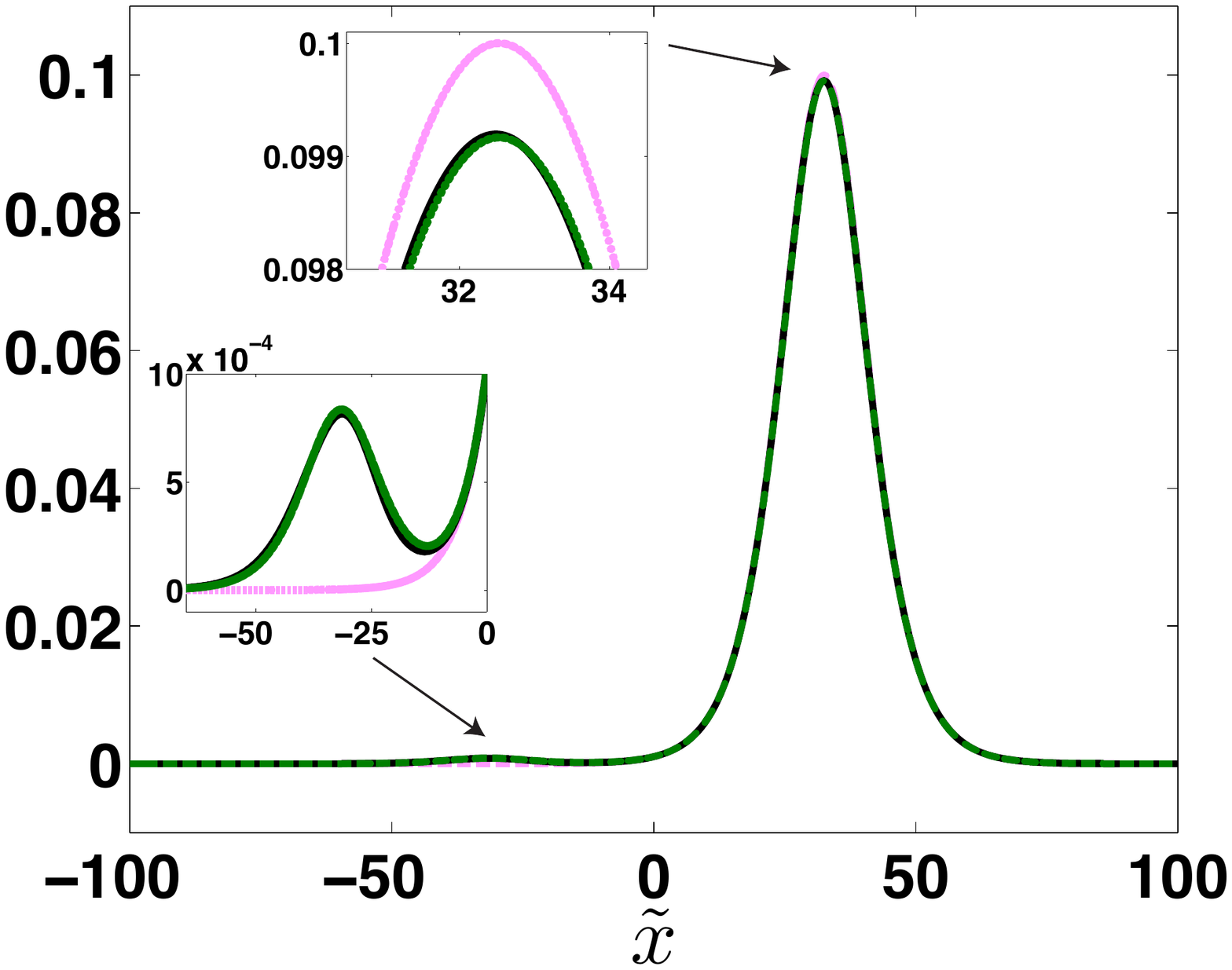} &                       
\includegraphics[width=2.45in]{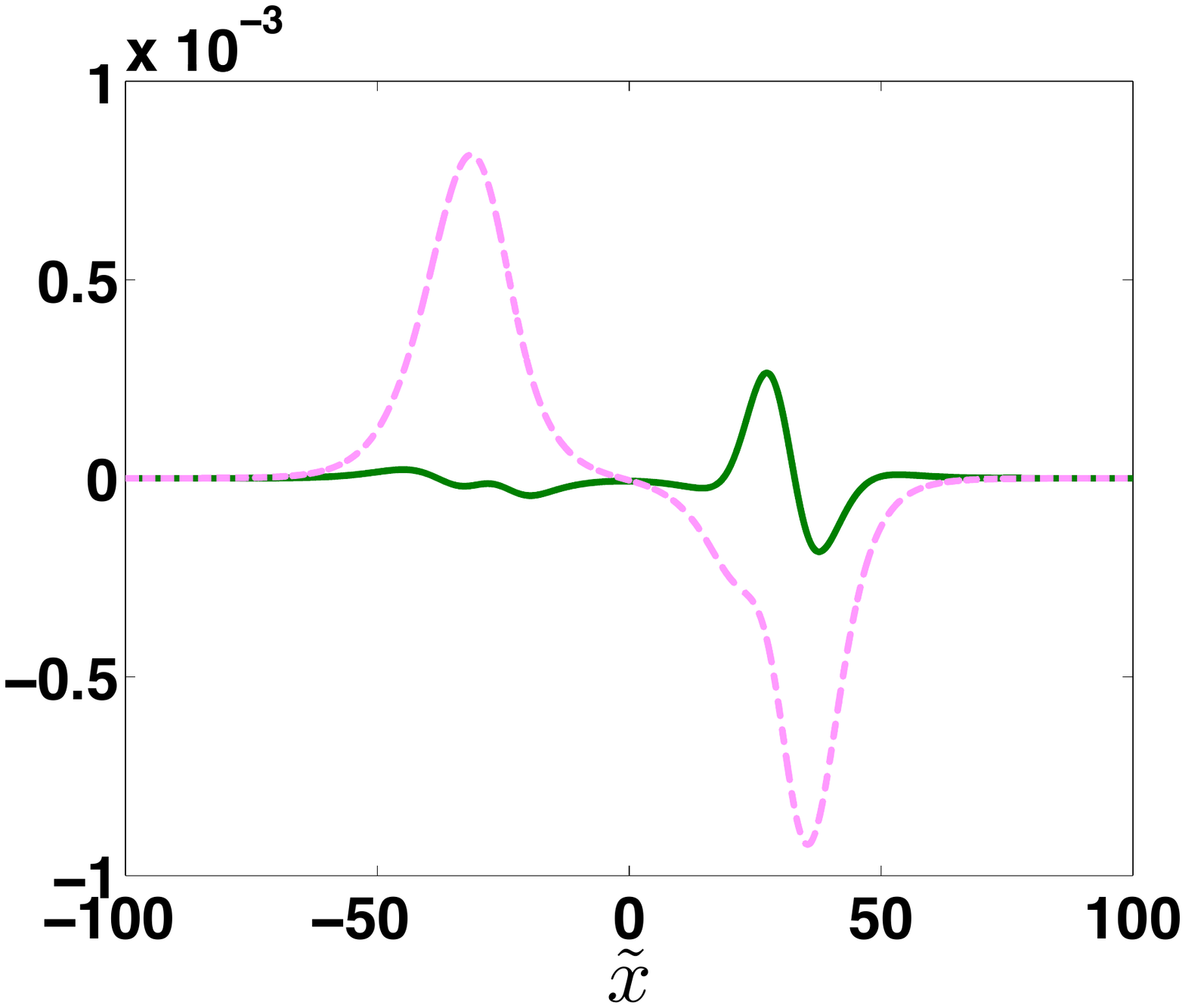}  &       
\includegraphics[width=2.45in]{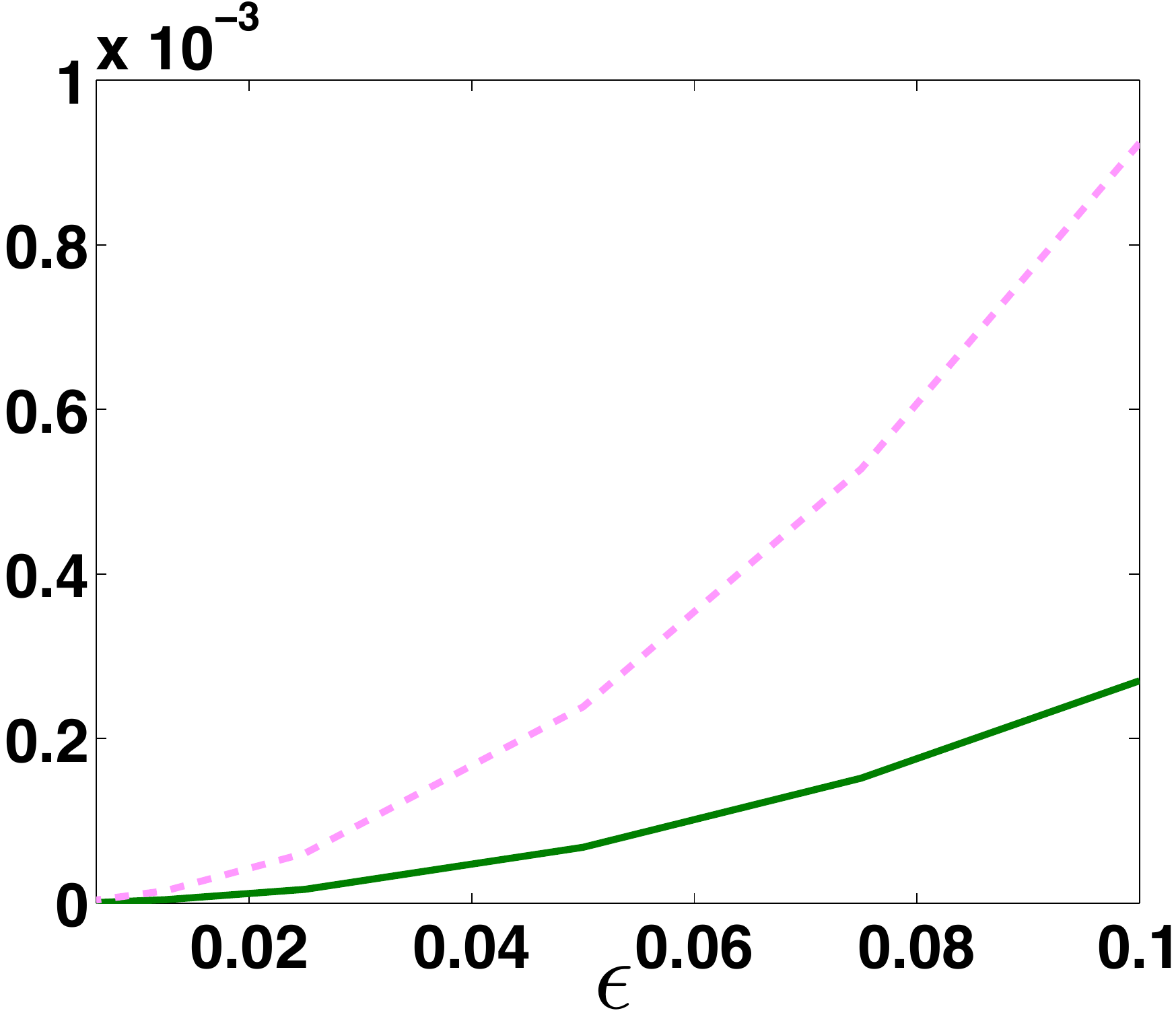} \\                                     
\quad  \mbox{\footnotesize \bf (a) $f_{{\rm num}}$({\color{black} \bf ---}) \  $\tilde{f}_{2}$({\color{KMgreen} \bf - -}) \  $\tilde{f}_{1}$({\color{KMpink} \bf - -}) } &
\mbox{ \footnotesize \bf (b) \footnotesize  ($f_{{\rm num}}$  -  $\tilde{f}_{2}$)({\color{KMgreen} \bf ---})     \       ($f_{{\rm num}}$  -  $\tilde{f}_{1}$)({\color{KMpink} \bf - -}) }  & 
\quad  \mbox{\footnotesize \bf (c) $e^2_{{\rm t_a}}$({\color{KMgreen} \bf ---}) \  $e^1_{{\rm t_a}}$({\color{KMpink} \bf - -})  }   
\end{array}$
\end{center}
\caption{\small Right-propagating 1-soliton initial conditions. (a) Numerical solution compared with the weakly nonlinear solution $\tilde{f}_2$ and $\tilde{f}_1$, and (b) the respective absolute errors, both for $\epsilon=0.1$ and $\tilde{t}=32\ (\approx t_a)$. (c) Maximum absolute errors  $e^2_{t_a}$ and $e^1_{t_a}$ versus $\epsilon$ at $\tilde{t}\approx t_a$. All other parameters are $k=\frac{1}{\sqrt{3}}$ and $\alpha=0$.}  
\label{figure:1sol_1way_plots_error}
\end{figure}

Figure \ref{figure:1sol_1way_3D} depicts the time evolution of the weakly nonlinear solution (\ref{eqn:ex1_WNS_num}).
Figure \ref{figure:1sol_1way_plots_error}a and \ref{figure:1sol_1way_plots_error}b highlight the difference in the numerical solution compared with the weakly nonlinear solution, taken up to leading and second order, for a particular time and choice of $\epsilon$. The difference between solution (\ref{eqn:ex1_WNS_num}) and the numerical solution, in Fig. \ref{figure:1sol_1way_plots_error}a, is almost indistinguishable, whilst a considerable difference can be observed for the leading order solution. 

Figure \ref{figure:1sol_1way_plots_error}c displays the maximum absolute error for varying $\epsilon$ at the corresponding time $\tilde{t}\approx t_a$. As expected the errors reduce as $\epsilon \rightarrow 0$, and for the weakly nonlinear solution (\ref{eqn:ex1_WNS_num}), the maximum of the errors remains within its expected accuracy of $O(\epsilon^3)$.
Each of the errors plotted in Fig. \ref{figure:1sol_1way_plots_error}c are greater in magnitude than the corresponding errors of the numerical scheme for the same parameters (but for different initial conditions), given in table (\ref{table:numerics_error_eps}) of Section \ref{sec:numerics}. Therefore we believe these plots provide a true measure of the accuracy of the weakly nonlinear solution and are not accountable for some numerical artefact of the finite difference scheme.
%
%
%

\subsection{Right- and left-propagating 1-soliton initial conditions}
The case of initial conditions of the IVP (\ref{eqn:IVP}) in the form of a right- and left-propagating single KdV soliton solution, yields the leading order solutions
\begin{eqnarray*}
f^+&=& f^+_1(\eta,T) = 12 \frac{\partial^2}{\partial \eta^2} {\rm log}(1+e^{\theta(\eta,T)}), \qquad
f^-  = f^-_1(\xi,T)  = 12 \frac{\partial^2}{\partial \xi^2} {\rm log}(1+e^{\theta(\xi,-T)}).
\end{eqnarray*}
From (\ref{eqn:nsol_both_ICs}) this corresponds to the initial conditions of the IVP (\ref{eqn:IVP}) in the form
\begin{eqnarray}
f|_{t=0}= 24 \frac{\partial^2}{\partial x^2} {\rm log} (1+e^{\theta(x,0)}), \qquad 
f_t|_{t=0} = 0.
\label{eqn:1_sol_1way_ICs}
\end{eqnarray}
The weakly nonlinear solution is therefore in the form (\ref{eqn:nsol_both_WNS_uv}) where in this particular case $U(x,T)= {\rm log} (1+e^{\theta(x,-T)})$ and $V(x,T)= {\rm log} (1+e^{\theta(x,T)})$. Explicitly evaluating each of the terms in the weakly nonlinear solution yields the solution in the following form
\begin{eqnarray}
f&=& 3k^2\left[{\rm sech}^2\theta^{\xi -} + {\rm sech}^2\theta^{\eta +} \right] 
+  \frac{9k^4\epsilon}{2}\left\lbrace  - \frac 1 6 \left[{\rm sech}^2 \theta^{x +} - {\rm sech}^2 \theta^{x -}\right]^{x=\eta}_{x=\xi}  
\right.\nonumber \\
&& \left. \left. -  {\rm sech}^2\theta^{\xi -} \left[ \frac 1 2 \left({\rm sech}^2\theta^{\eta +}   +  {\rm sech}^2\theta^{\xi +}  \right) 
\right. \right.        - {\rm tanh}\ \theta^{\xi -} \left ( {\rm tanh}\ \theta^{\eta +}   +   {\rm tanh}\ \theta^{\xi +}  \right)\right]
 \nonumber \\
&& \left. \left.
-  {\rm sech}^2\theta^{\eta +} \left[ \frac 1 2 \left({\rm sech}^2\theta^{\xi -}   +  {\rm sech}^2\theta^{\eta -}  \right) \right.
  - {\rm tanh}\ \theta^{\eta +} \left ( {\rm tanh}\ \theta^{\xi -}   +   {\rm tanh}\ \theta^{\eta -}  \right)\right]
  \right\rbrace+ O(\epsilon^2),
  \label{eqn:ex2_WNS}
\end{eqnarray} 
where we use the notation  $  \theta^{x \pm} = \frac 1 2 \theta(x, \pm T)$.

We next investigate the error of the weakly nonlinear solution by firstly transforming the variables in (\ref{eqn:ex2_WNS}) to the same form used in the numerics, and for numerical simulations we use the following initial conditions to coincide with (\ref{eqn:1_sol_1way_ICs})
\begin{eqnarray*}
 f_{i,0}=6 k^2 \epsilon\ {\rm sech}^2 \left(\frac{\sqrt{\epsilon} k \tilde{x} + \alpha}{2}\right),
 \qquad  f_{i,1}=3k^2 \epsilon\left[{\rm sech}^2 \left(\frac{\sqrt{\epsilon} k (\tilde{x}-\kappa)+\alpha}{2}\right) + {\rm sech}^2\left(\frac{\sqrt{\epsilon} k (\tilde{x} + \kappa) + \alpha}{2} \right)  \right].
\end{eqnarray*}  
We again choose some suitable $k$ to ensure the validity of the weakly nonlinear solution.

\begin{figure}[htbp]
\begin{center}
\includegraphics[width=10cm]{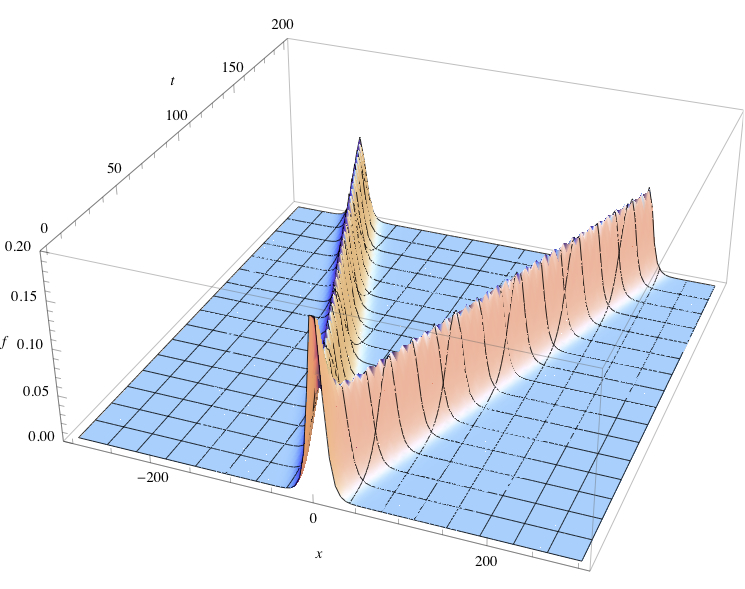}
\caption{\small Evolution of the weakly nonlinear solution for right- and left-propagating 1-soliton initial conditions, with $\epsilon=0.1,\ k=\frac{1}{\sqrt{3}}$ and $\alpha=0$.}
\label{figure:1sol_2way_3D}
\end{center}
\end{figure}

\begin{figure}[htbp]
\begin{center}$
\begin{array}{ccc}
\includegraphics[width=2.45in]{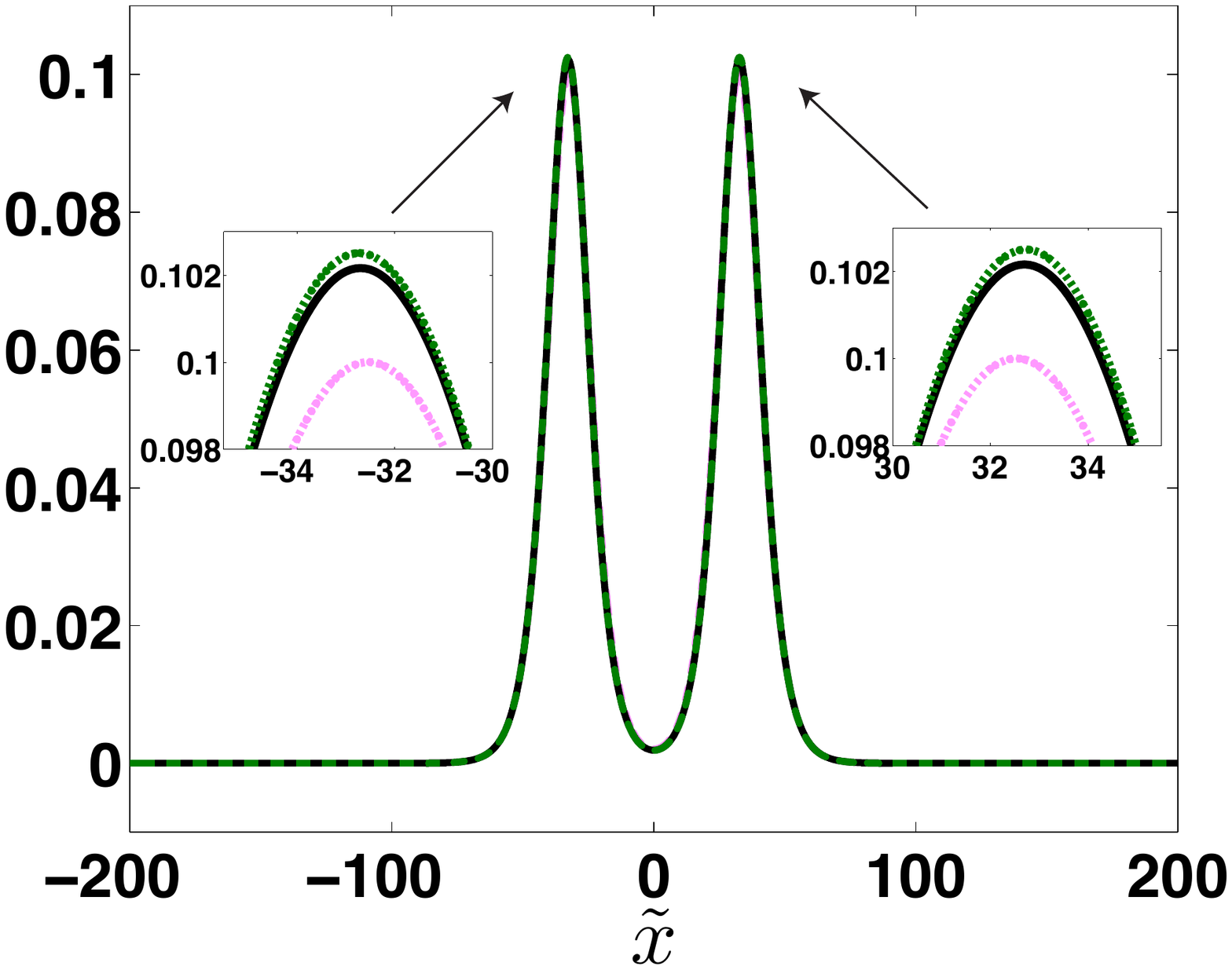} &                       
\includegraphics[width=2.40in]{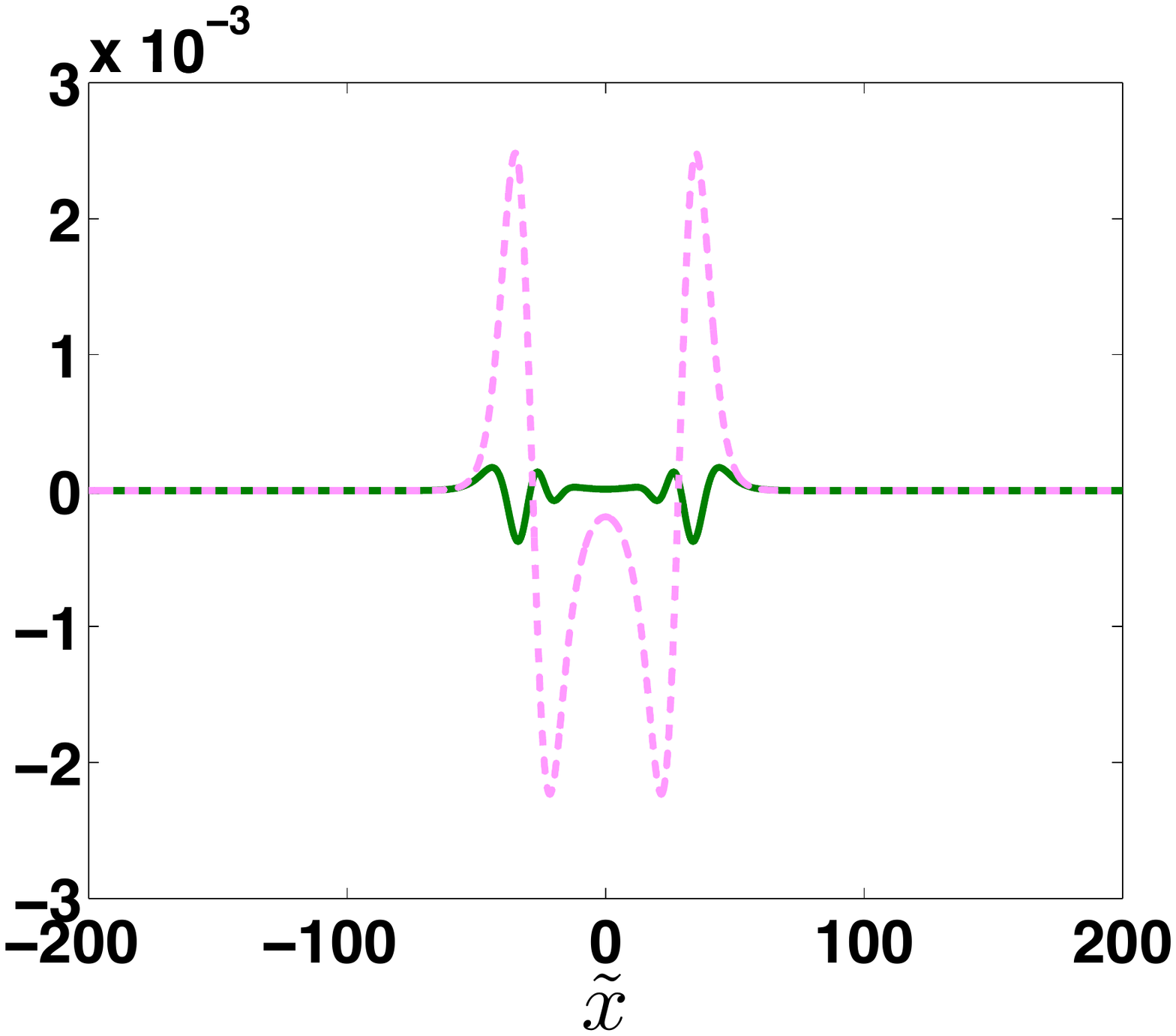} &			
\includegraphics[width=2.4in]{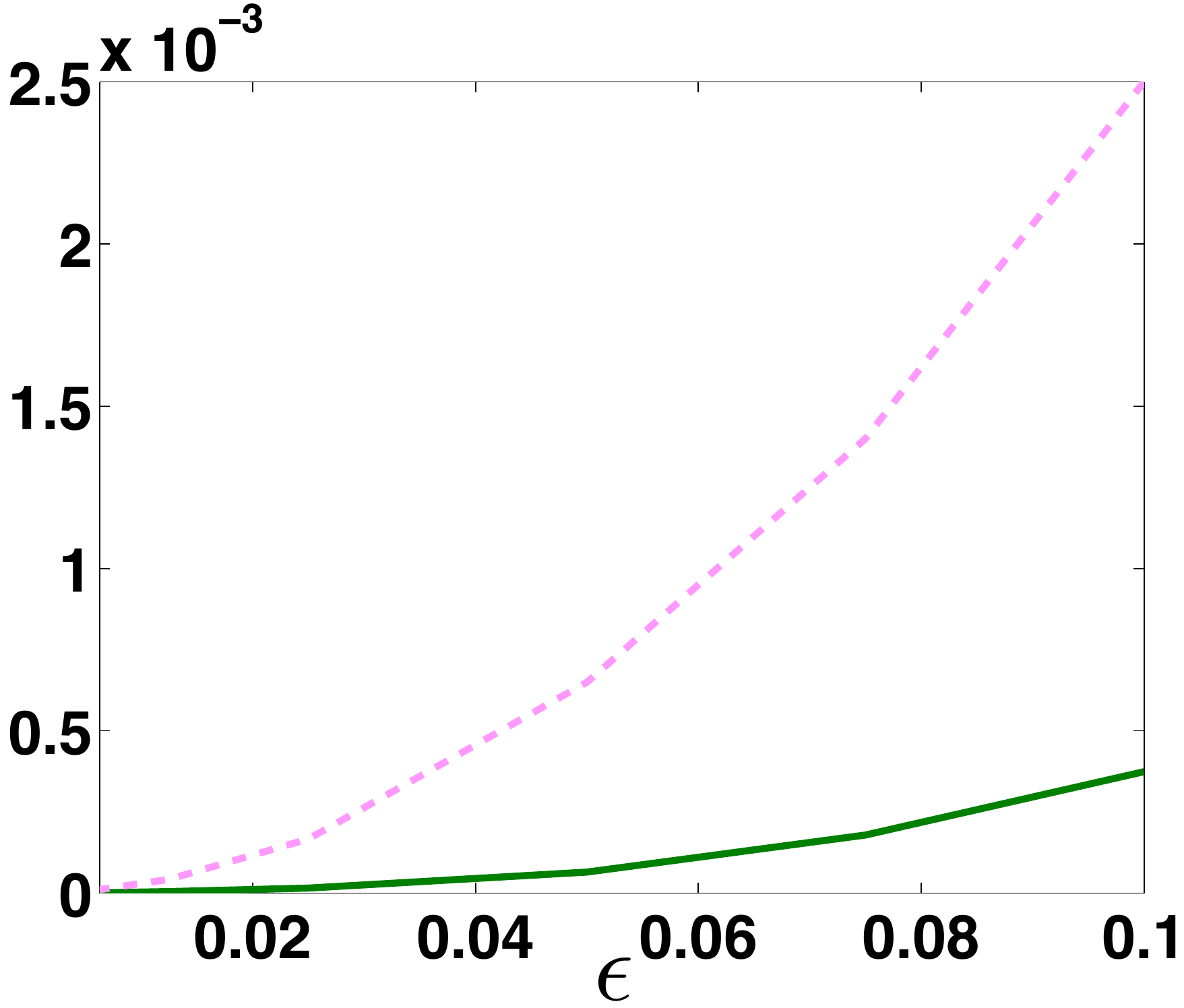} \\ 		
\quad  \mbox{\footnotesize \bf (a) $f_{{\rm num}}$({\color{black} \bf ---}) \  $\tilde{f}_{2}$({\color{KMgreen} \bf - -}) \  $\tilde{f}_{1}$({\color{KMpink} \bf - -}) } &
\mbox{ \footnotesize \bf (b) \footnotesize  ($f_{{\rm num}}$  -  $\tilde{f}_{2}$)({\color{KMgreen} \bf ---})     \       ($f_{{\rm num}}$  -  $\tilde{f}_{1}$)({\color{KMpink} \bf - -}) }  & 
\quad  \mbox{\footnotesize \bf (c) $e^2_{{\rm t_a}}$({\color{KMgreen} \bf ---}) \  $e^1_{{\rm t_a}}$({\color{KMpink} \bf - -})  }   
\end{array}$
\end{center}
\caption{\small Right- and left-propagating 1-soliton initial conditions. (a) Numerical solution compared with the weakly nonlinear solution $\tilde{f}_2$ and $\tilde{f}_1$, and (b) the respective absolute errors, both for $\epsilon=0.1$ and $\tilde{t}=32\ (\approx t_a)$. (c) Maximum absolute errors  $e^2_{t_a}$ and $e^1_{t_a}$ versus $\epsilon$ at $\tilde{t}\approx t_a$. All other parameters are $k=\frac{1}{\sqrt{3}}$ and $\alpha=0$.}  
\label{figure:1sol_2way_3plots}
\end{figure}

Figure \ref{figure:1sol_2way_3D} depicts the evolution over time of the weakly nonlinear solution in the same form used in the numerics. Figure \ref{figure:1sol_2way_3plots}a and \ref{figure:1sol_2way_3plots}b display the behaviour of the numerical solution compared with the weakly nonlinear solution, up to leading and second order, for fixed $\tilde{t}$ and $\epsilon$. Figure \ref{figure:1sol_2way_3plots}c displays the maximum of the absolute errors for the weakly nonlinear solution taken up to each order, for various $\epsilon$ and the corresponding $\tilde{t}\approx t_a$. 
Similar to the previous example the weakly nonlinear solution derived for this example is as accurate as expected ($O(\epsilon^3)$), and the accuracy improves as $\epsilon\rightarrow 0$. Again it's clear that there is a significant reduction in the maximum absolute error when the second order terms are included within the weakly nonlinear solution. 
%
%
%
%
%
%
%
%
%
%
%
%
%

\newpage
\subsection{Right-propagating 2-soliton initial conditions}
We next examine another particular example of the weakly nonlinear solution derived in Section \ref{sec:1way_nsol}, namely initial conditions of the IVP (\ref{eqn:IVP}) in the form of a right-propagating  2-soliton solution of the KdV equation. Therefore we have the following leading order solutions
\begin{eqnarray*}
f^+=0, \qquad f^-  = f^-_2(\xi,T)  =  12 \frac{\partial^2}{\partial \xi^2} {\rm log}\left(1+e^{\theta_1(\xi,-T)}+e^{\theta_2(\xi,-T)} + Ce^{\theta_1(\xi,-T)+\theta_2(\xi,-T)}\right) ,
\end{eqnarray*}
where $C=[(k_1-k_2)/(k_1+k_2)]^2$ and $\theta_i(x,T)=k_ix + \frac{k^3_i}{2}T + \alpha_i$ for $i=1,2$. From  (\ref{eqn:one_way_N_sol_ICs}) the corresponding initial conditions of the IVP (\ref{eqn:IVP}) are in the form
\begin{eqnarray}
f|_{t=0}= 12 \frac{\partial^2}{\partial x^2} {\rm log} \left(1 + e^{\theta_1(x,0)}+e^{\theta_{2}(x,0)} + Ce^{\theta_{1}(x,0)+\theta_{2}(x,0)}\right), \quad 
f_t|_{t=0} = - \frac{\partial}{\partial x}f|_{t=0},
\label{eqn:2sol_1way_ICs}
\end{eqnarray}
and from (\ref{eqn:WNS_1way}) the weakly nonlinear solution for this example can be explicitly expressed as

\begin{eqnarray}
f = 6 \frac{(k_1-k_2)^2 + \sqrt{C}\left(k_1^2 {\rm \ cosh}\ \theta_2^{\xi -} + k_2^2 {\rm \ cosh}\ \theta_1^{\xi -}\right)}
{\left[ {\rm \ cosh} \left(\frac{\theta_1^{\xi -} - \theta_2^{\xi -}}{2}\right) + \sqrt{C} {\rm \ cosh} \left(\frac{\theta_1^{\xi -} + \theta_2^{\xi -}}{2}\right)\right]^2 } 
+\frac{3\epsilon}{2} \left\lbrace  \frac{D + \sqrt{C}\left(k_1^4 {\rm \ cosh}\ \theta_2^{\eta -} + k_2^4 {\rm \ cosh}\ \theta_1^{\eta -}\right)}
{\left[ {\rm \ cosh} \left(\frac{\theta_1^{\eta -} - \theta_2^{\eta -}}{2}\right) + \sqrt{C} {\rm \ cosh} \left(\frac{\theta_1^{\eta -} + \theta_2^{\eta -}}{2}\right)\right]^2 }  \right.
\nonumber \\
 - \left.\frac{D + \sqrt{C}\left(k_1^4 {\rm \ cosh}\ \theta_2^{\xi -} + k_2^4 {\rm \ cosh}\ \theta_1^{\xi -}\right)}
{\left[ {\rm \ cosh} \left(\frac{\theta_1^{\xi -} - \theta_2^{\xi -}}{2}\right) + \sqrt{C} {\rm \ cosh} \left(\frac{\theta_1^{\xi -} + \theta_2^{\xi -}}{2}\right)\right]^2 }  \right\rbrace + O(\epsilon^2). \hspace{7cm}
\label{eqn:ex3_WNS_trig}
\end{eqnarray}
where $D=(k_1-k_2)^2(k_1^2+k_2^2)$ and we use the notation $\theta^{x\pm}_i =k_ix \pm \frac{k_i^3}{2} T + \hat{\alpha}_i$ (where we shift $\alpha_i \rightarrow \hat{\alpha}_i - {\rm ln}\sqrt{C}$) for $i=1,2$. We next examine the error of (\ref{eqn:ex3_WNS_trig}) by writing it in the form used for the numerics and impose the following initial conditions for numerical simulations
\begin{eqnarray*}
f_{i,0} = 6\epsilon \frac{(k_1-k_2)^2 + \sqrt{C}\left(k_1^2 {\rm \ cosh}\ \tilde{\theta}_{2}^{x0} + k_2^2 {\rm \ cosh}\ \tilde{\theta}_{1}^{x0}\right)}
{\left[ {\rm \ cosh} \left(\frac{\tilde{\theta}_{1}^{x0} - \tilde{\theta}_{2}^{x0}}{2}\right) + \sqrt{C} {\rm \ cosh} \left(\frac{\tilde{\theta}_{1}^{x0} + \tilde{\theta}_{2}^{x0}}{2}\right)\right]^2 }, \quad
\qquad f_{i,1} = f_{i,0}|_{\tilde{x}=\tilde{x}-\kappa},
\end{eqnarray*}
where $\tilde{\theta}^{x0}_{i}=k_i\tilde{x}\sqrt{\epsilon} + \hat{\alpha}_i$ for $i=1,2$, and we choose some $k_1$ and $k_2$ appropriately, to ensure applicability of the weakly nonlinear solution.

\begin{figure}[h]
\begin{center}$
\begin{array}{ccc}
\quad \tilde{t}=0 & \quad \tilde{t}=32&  \quad \tilde{t}=300 \\
&&\\
\includegraphics[width=2.1in]{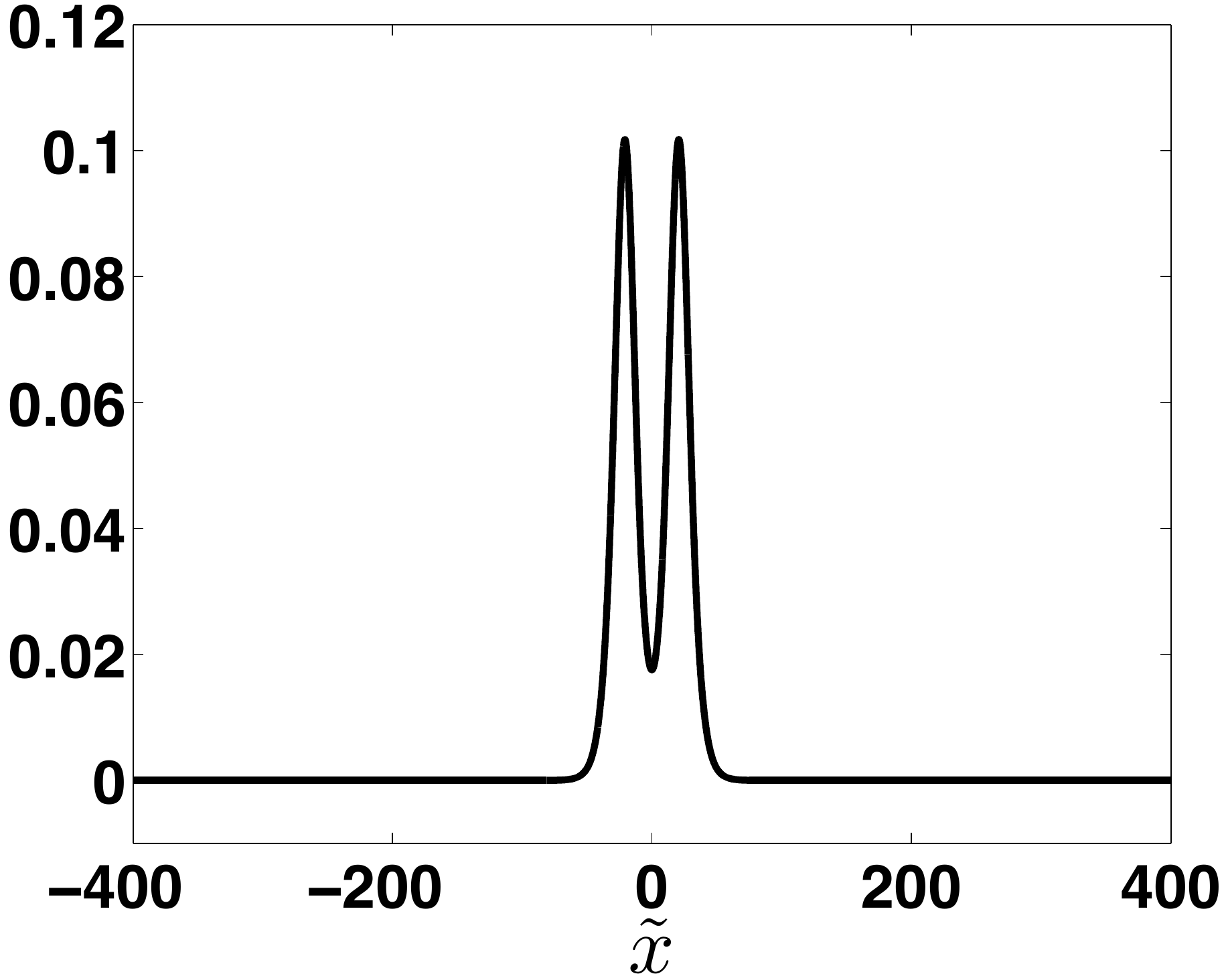} &  
\includegraphics[width=2.1in]{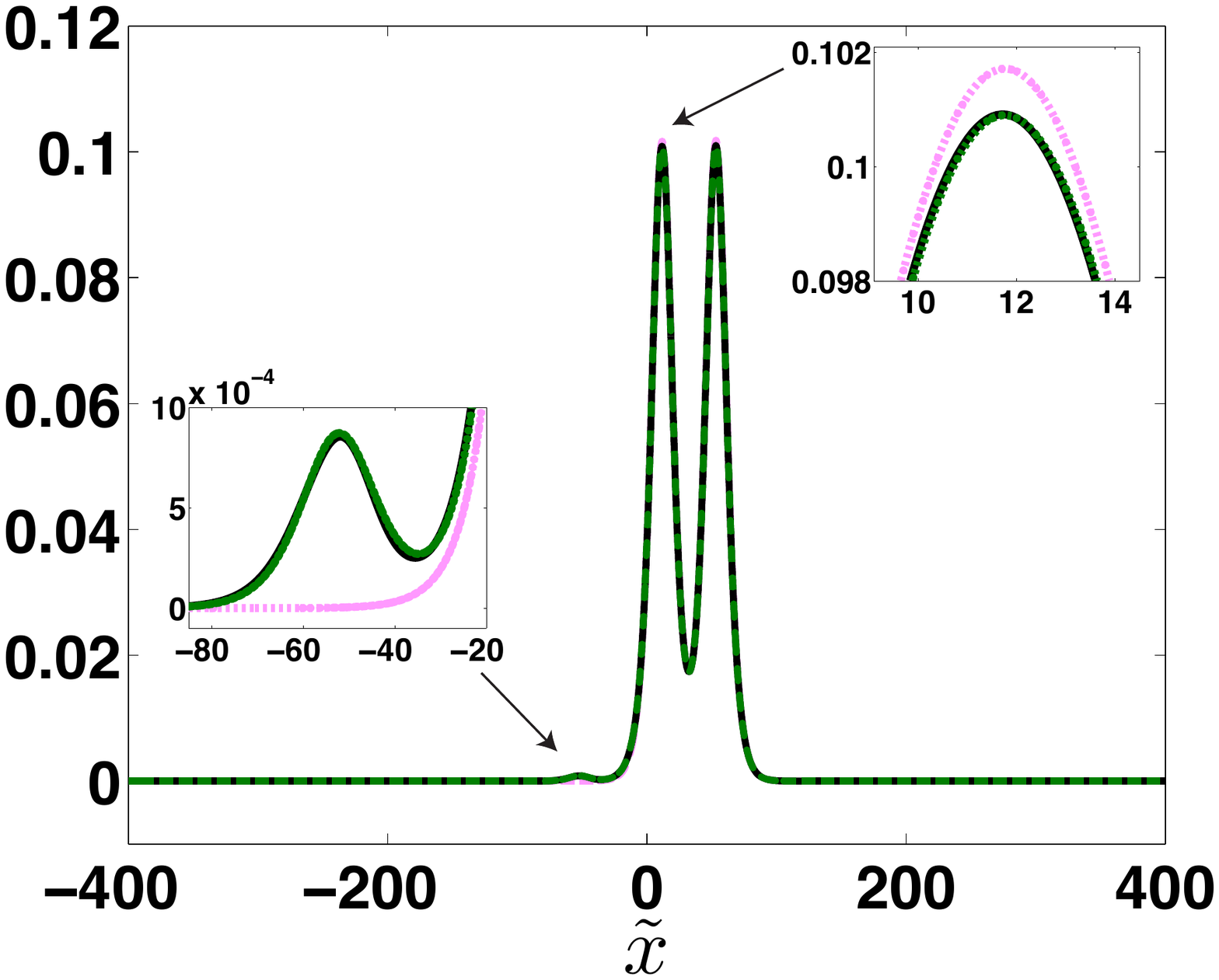} &       
\includegraphics[width=2.1in]{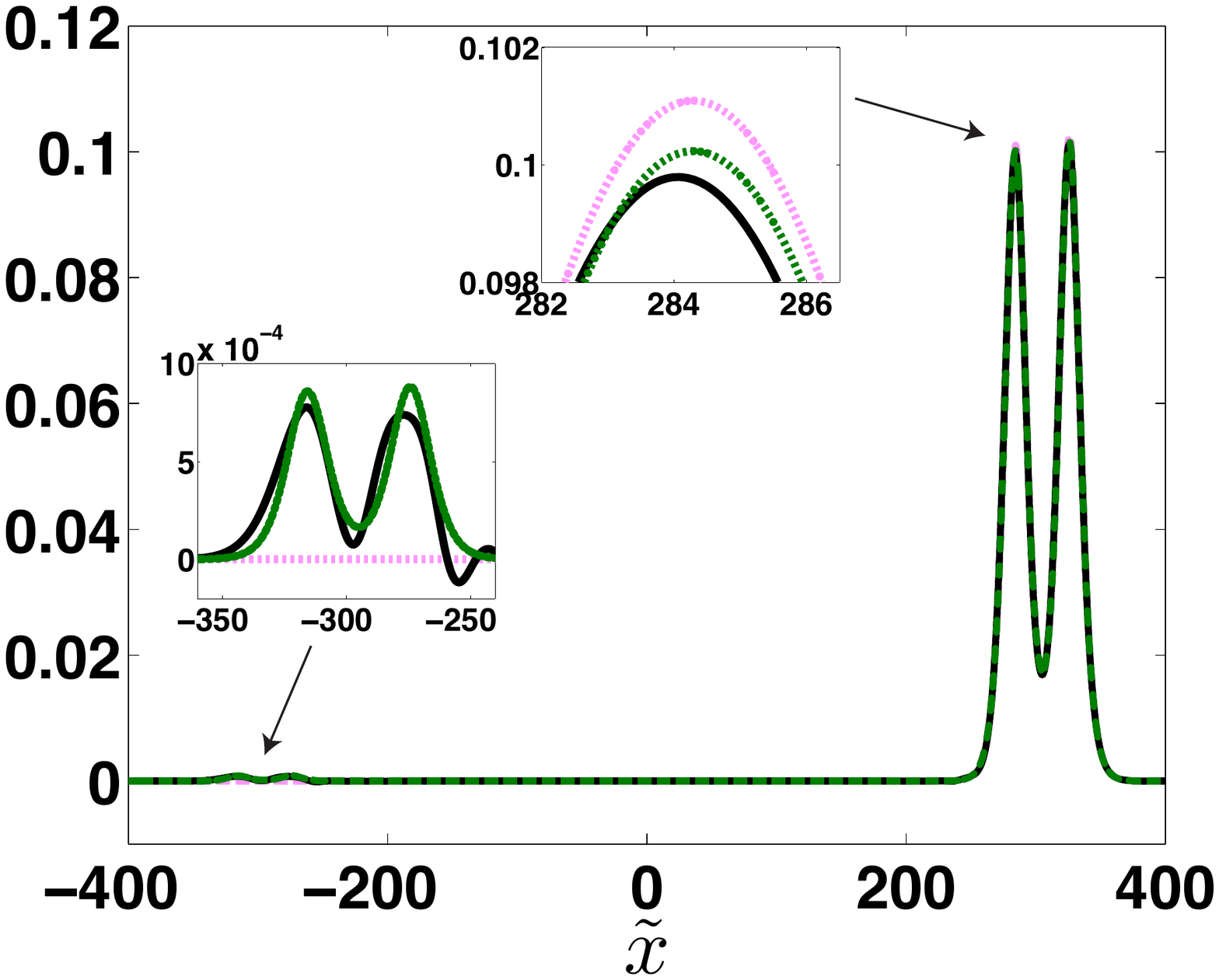} \\   
\quad  \mbox{\footnotesize \bf (a) $f_{{\rm num}}, \tilde{f}_{2},   \tilde{f}_{1}$({\color{black} \bf ---})  }  &
\quad  \mbox{\footnotesize \bf (b) $f_{{\rm num}}$({\color{black} \bf ---}) $\tilde{f}_{2}$({\color{KMgreen} \bf - -}) $\tilde{f}_{1}$({\color{KMpink} \bf - -}) } & 
\quad  \mbox{\footnotesize \bf (c) $f_{{\rm num}}$({\color{black} \bf ---}) $\tilde{f}_{2}$({\color{KMgreen} \bf - -}) $\tilde{f}_{1}$({\color{KMpink} \bf - -}) }         \\\\
\includegraphics[height=1.7in,width=2.3in]{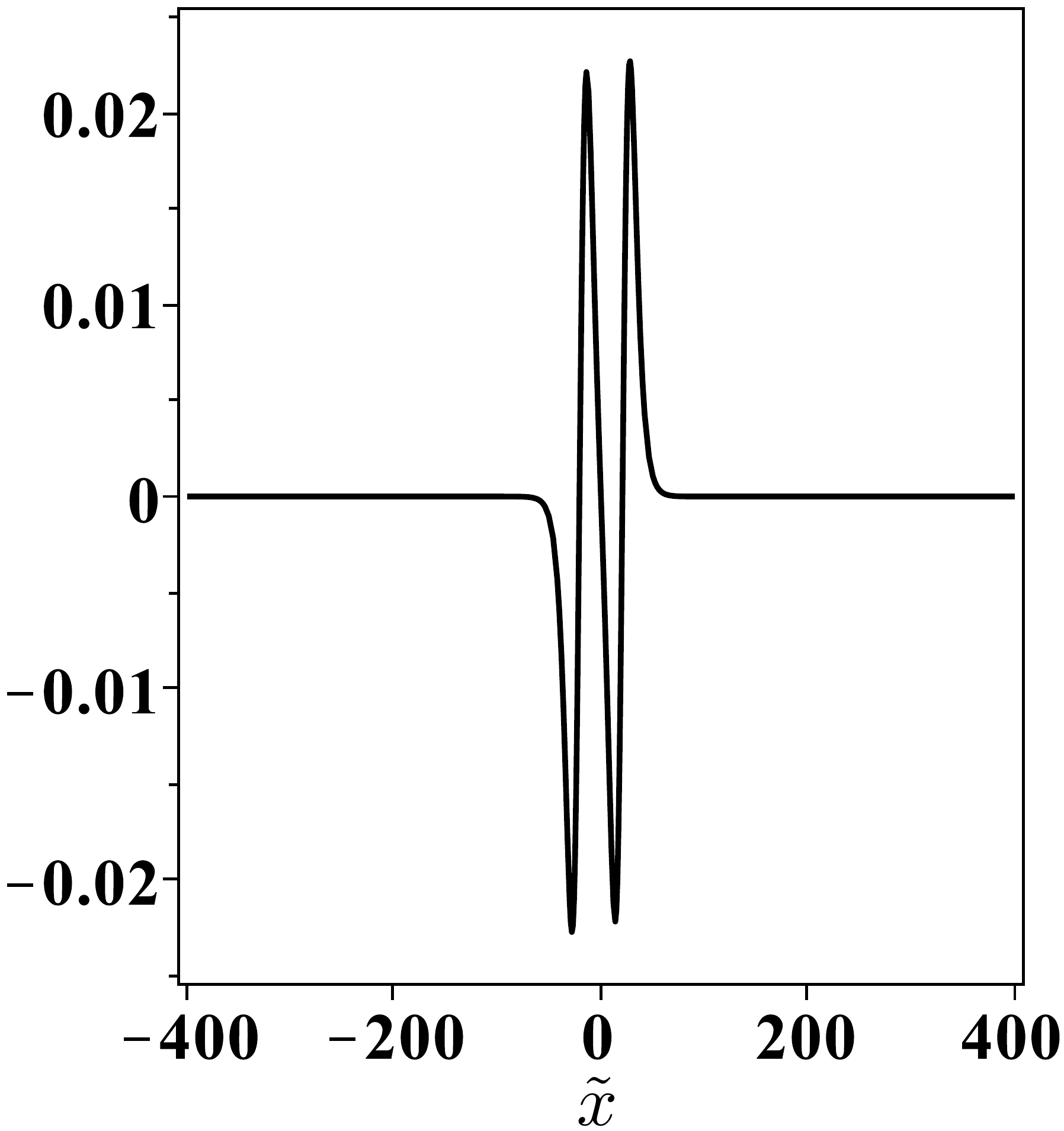}  &                                         
\includegraphics[width=2.1in]{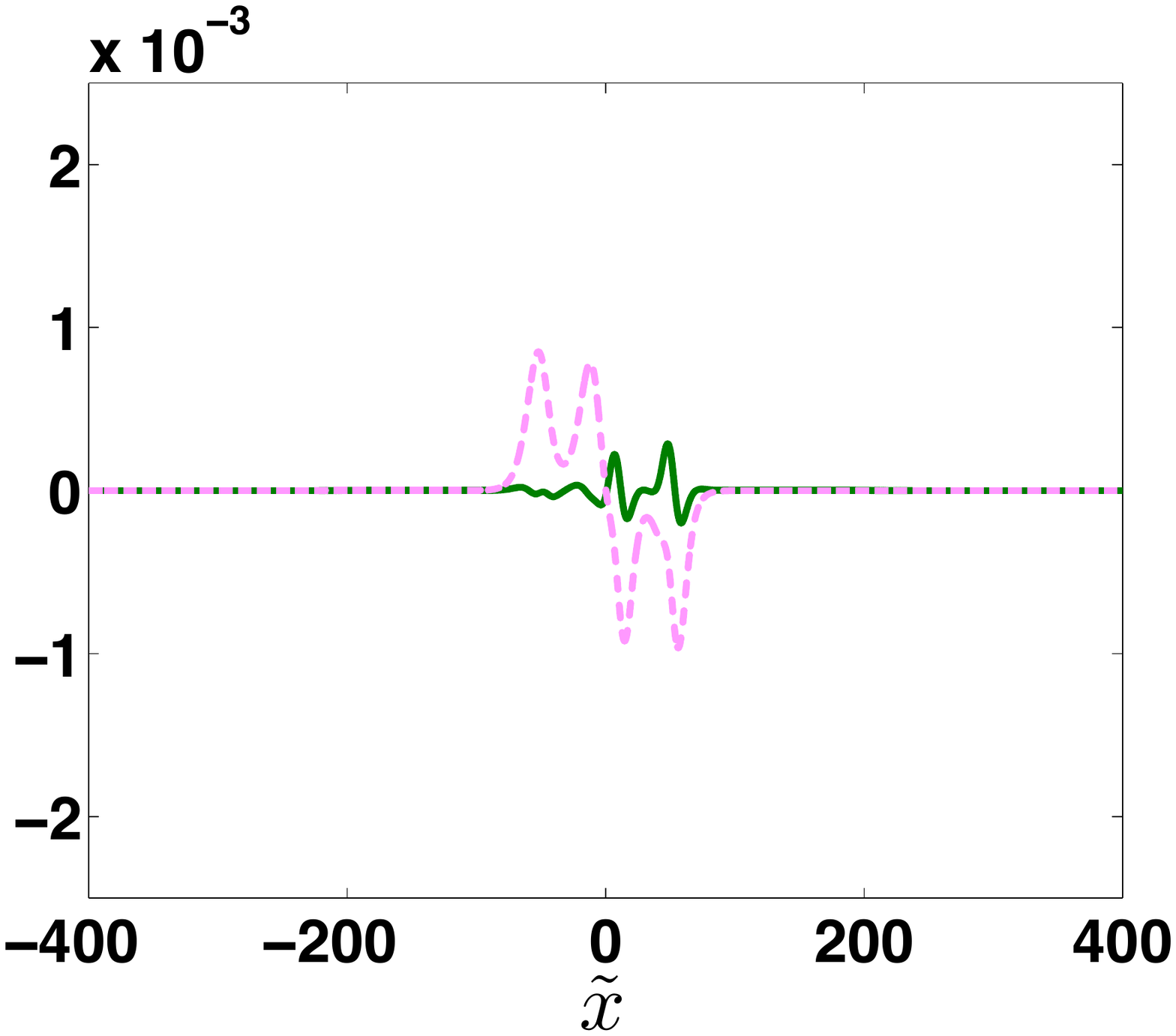} &                                                   
\includegraphics[width=2.1in]{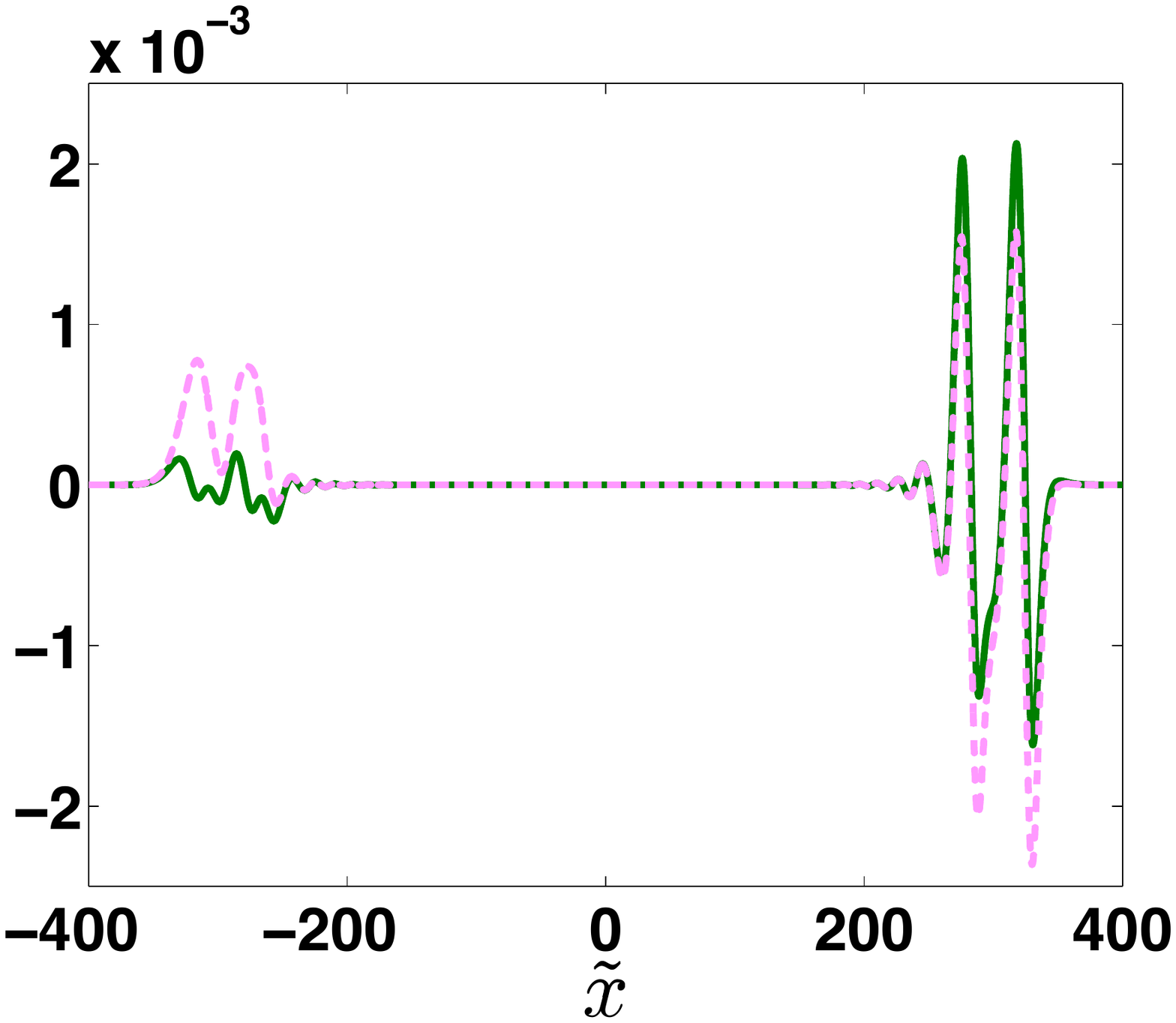}  \\           
\quad  \mbox{\footnotesize \bf (d) $\frac{\partial}{\partial t} (f_{{\rm num}}, \tilde{f}_{2},   \tilde{f}_{1})$({\color{black} \bf ---})  } &
\mbox{ \footnotesize \bf (e) \scriptsize  ($f_{{\rm num}}$  -  $\tilde{f}_{2}$)({\color{KMgreen} \bf ---})     \       ($f_{{\rm num}}$  -  $\tilde{f}_{1}$)({\color{KMpink} \bf - -}) }  & 
 \mbox{ \footnotesize \bf (f) \scriptsize  ($f_{{\rm num}}$  -  $\tilde{f}_{2}$)({\color{KMgreen} \bf ---})     \       ($f_{{\rm num}}$  -  $\tilde{f}_{1}$)({\color{KMpink} \bf - -}) }    \\
\end{array}$
\end{center}
\caption{\small Right-propagating 2-soliton initial conditions (a) $\&$ (d). Evolution of the numerical solution compared with the weakly nonlinear solution $\tilde{f_2}$ and $\tilde{f_1}$ at (b) $\tilde{t}=32\ (\approx t_a)$, (c) $\tilde{t}=300\ (\approx t_b)$, and the absolute errors at each of the respective times (e) $\&$ (f). All other parameters are $\epsilon=0.1, k_1=0.61, k_2=0.56$ and $\hat{\alpha}_1=\hat{\alpha}_2=0$.} 
\label{figure:2sol_1way}
\end{figure}

In Fig. \ref{figure:2sol_1way} we consider the case where the amplitudes of the leading order parts of the solution are close ($k_1=0.61, k_2=0.56$) and the phase shifts $\hat{\alpha}_1=\hat{\alpha}_2=0$, such that their relative positions are initially close together. 
The initial conditions of the IVP (\ref{eqn:IVP}) for this example are shown in Fig.  \ref{figure:2sol_1way}a and  \ref{figure:2sol_1way}d, and the evolution of the weakly nonlinear solution, up to first and second order, and the numerical solution, are displayed in Fig. \ref{figure:2sol_1way}b and \ref{figure:2sol_1way}c, all for times within the region of validity of the weakly nonlinear solution.  The corresponding absolute errors up to each order are displayed in Fig. \ref{figure:2sol_1way}e and \ref{figure:2sol_1way}f at each of the respective times. 
It's clear that the accuracy of the weakly nonlinear solution is within the expected accuracy of $O(\epsilon^3)$ throughout the time interval considered. There is a noticeable improvement in the coincidence of the numerical solution with the weakly nonlinear solution taken up to second order compared to the leading order solution, best emulated at earlier time (i.e. in Fig. \ref{figure:2sol_1way}b and \ref{figure:2sol_1way}e). 

We also examined each combination for when the two leading order components of the weakly nonlinear solution are initially close or separate and when the amplitudes are similar or essentially different, controlled by the $\hat{\alpha}_i$'s and the $k_i$'s respectively. There were slight alterations to these results from the example presented in Fig. \ref{figure:2sol_1way} however the main features remained, that being $e^2_t$ remained within $O(\epsilon^3)$ throughout the same time interval and $e^2_t \rightarrow e^1_t$ as time increased beyond the region of the validity of the weakly nonlinear solution.

\subsection{Right- and left-propagating 2-soliton initial conditions}

Lastly in this section we consider another particular example of exactly solvable initial conditions, namely initial conditions of the IVP (\ref{eqn:IVP}) in the form of right and left-propagating 2-soliton solutions of the KdV equations. Therefore the leading order solutions are given by
\begin{eqnarray}
f^+&=& f^+_2(\eta,T)  \  = \ 12 \frac{\partial^2}{\partial \eta^2} {\rm log}\left(1+e^{\theta_1(\eta,T)}+e^{\theta_2(\eta,T)} + Ce^{\theta_1(\eta,T)+\theta_2(\eta,T)}\right) ,    \nonumber \\
f^-  &= &f^-_2(\xi,T) \ = \ 12 \frac{\partial^2}{\partial \xi^2} {\rm log}\left(1+e^{\theta_1(\xi,-T)}+e^{\theta_2(\xi,-T)} + Ce^{\theta_1(\xi,-T)+\theta_2(\xi,-T)}\right) ,
\label{eqn:2sol_2way_lead_terms}
\end{eqnarray}
and 
the initial conditions  (\ref{eqn:nsol_both_ICs}) for this example are in the form
\begin{eqnarray}
f|_{t=0}=  24  \frac{\partial^2}{\partial x^2} {\rm log} \left(1 + e^{\theta_1(x,0)}+e^{\theta_{2}(x,0)} + Ce^{\theta_{1}(x,0)+\theta_{2}(x,0)}\right), \quad 
f_t|_{t=0} = 0.
\label{eqn:2sol_2way_ICs}
\end{eqnarray}
The explicit form of the weakly nonlinear solution for this example is omitted here since the solution is rather lengthy, particularly due to the presence of third order derivatives of the log terms in (\ref{eqn:2sol_2way_lead_terms}). However the solution can be easily obtained from (\ref{eqn:nsol_both_WNS_uv}) using any computer algebra package (we used {\it Maple 14}). Indeed, the only operations required in determining (\ref{eqn:nsol_both_WNS_uv}) are differentiations. Thus,  once again we compare the weakly nonlinear solution with the numerical solution. Transforming the variables into the same form used in the numerics we use the following initial conditions for numerical simulations to coincide with (\ref{eqn:2sol_2way_ICs})
\begin{eqnarray*}
f_{i,0} = 12\epsilon \frac{(k_1-k_2)^2 + \sqrt{C}\left(k_1^2 {\rm \ cosh}\ \tilde{\theta}_{2}^{x0} + k_2^2 {\rm \ cosh}\ \tilde{\theta}_{1}^{x0}\right)}
{\left[ {\rm \ cosh} \left(\frac{\tilde{\theta}_{1}^{x0} - \tilde{\theta}_{2}^{x0}}{2}\right) + \sqrt{C} {\rm \ cosh} \left(\frac{\tilde{\theta}_{1}^{x0} + \tilde{\theta}_{2}^{x0}}{2}\right)\right]^2 } \quad
{\rm and} \quad f_{i,1} = \frac 1 2\left( f_{i,0}|_{\tilde{x}=\tilde{x}-\kappa}  +    f_{i,0}|_{\tilde{x}=\tilde{x}+\kappa}  \right),
\end{eqnarray*}
where we again choose the $k_i$'s appropriately in order to maintain the applicability of the weakly nonlinear solution.

\begin{figure}[htbp]
\begin{center}$
\begin{array}{ccc} 
\quad \tilde{t}=0 & \quad \tilde{t}=32&  \quad \tilde{t}=300 \\ \\
\includegraphics[height=1.81in,width=2.2in]{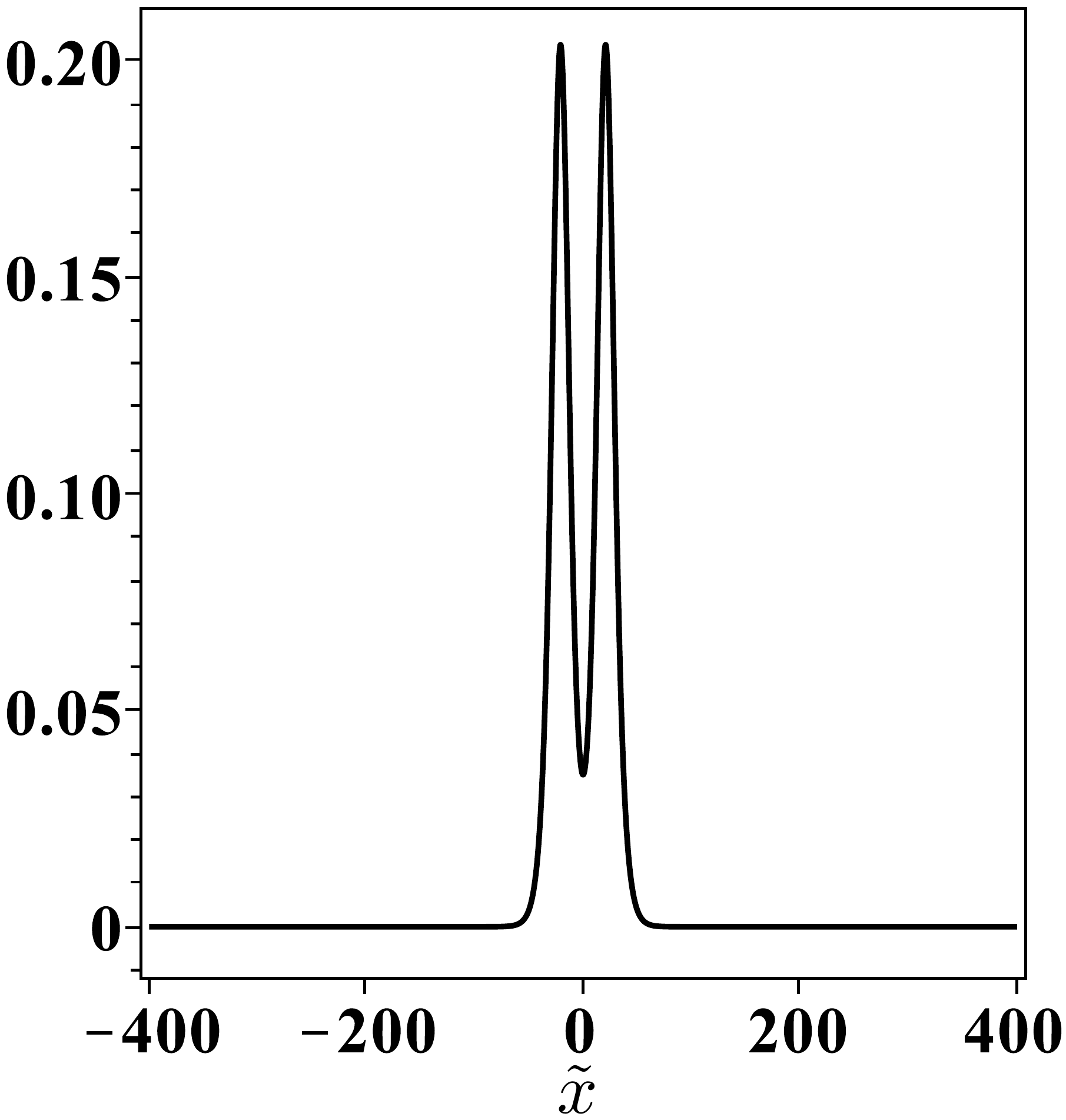} &   			
\includegraphics[height=1.8in,width=2.2in]{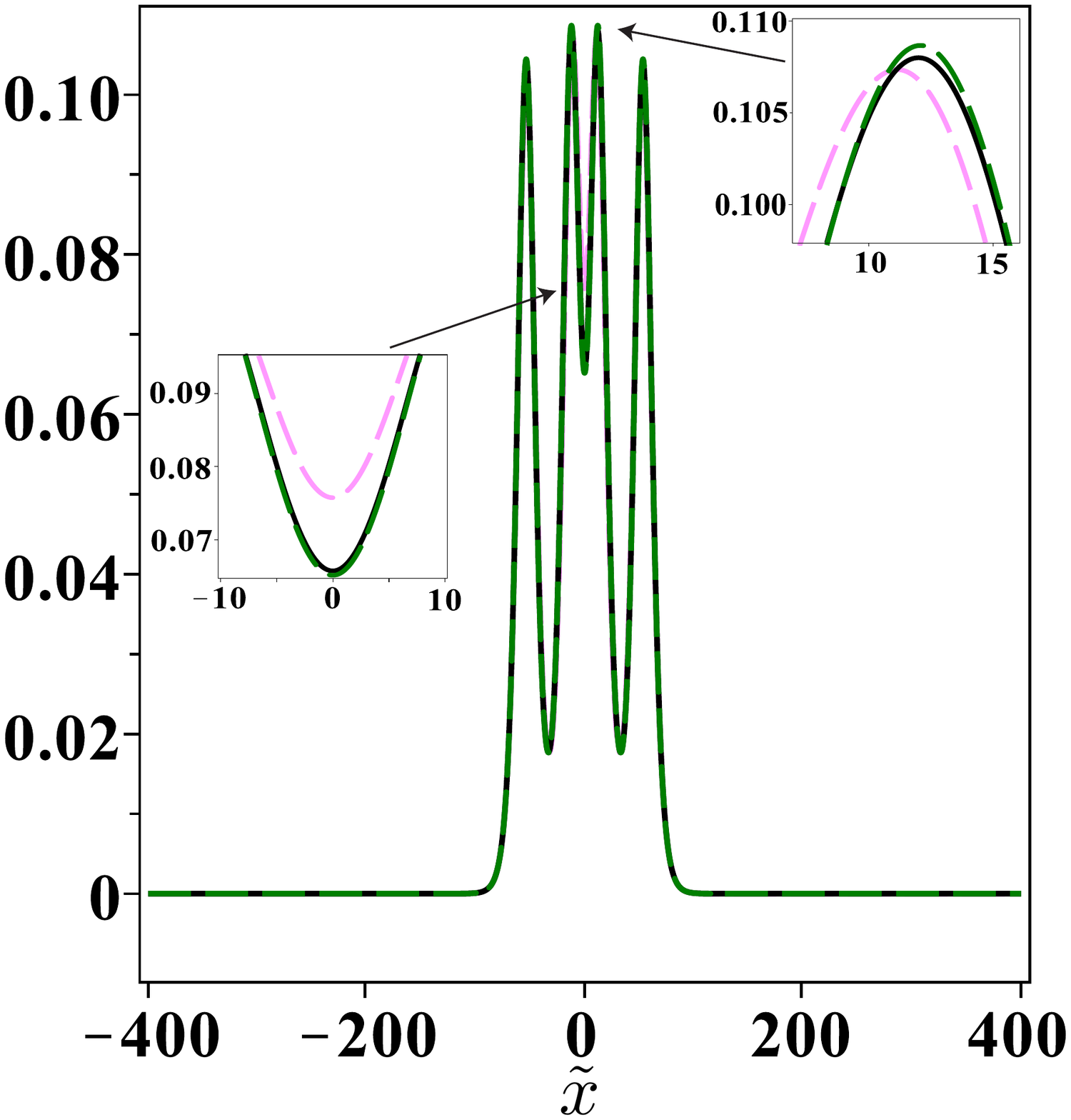} &
\includegraphics[height=1.8in,width=2.3in]{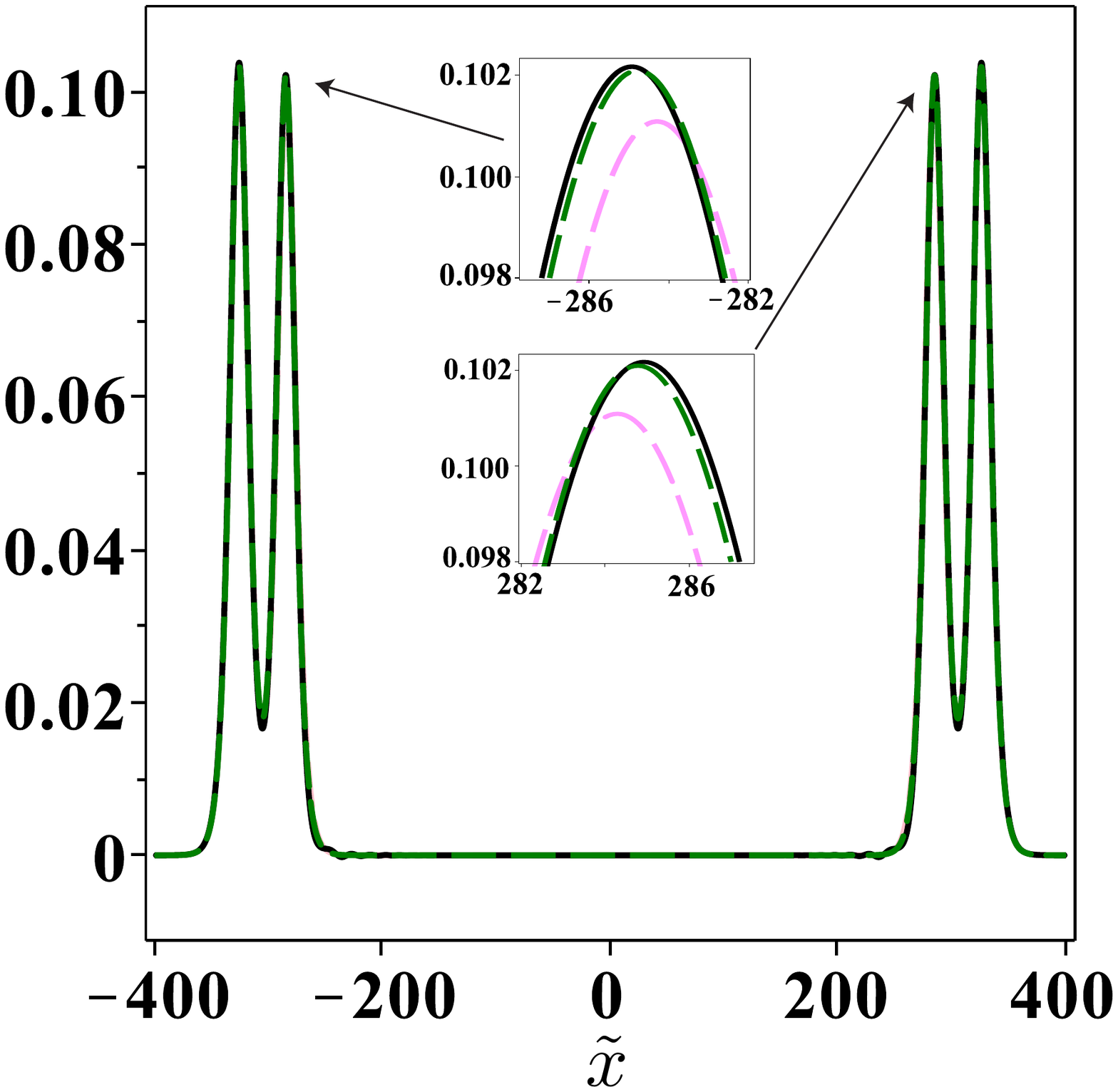} \\           		
\quad  \mbox{\footnotesize \bf (a) $f_{{\rm num}}, \tilde{f}_{2},   \tilde{f}_{1}$({\color{black} \bf ---})  }  &
\quad  \mbox{\footnotesize \bf (b) $f_{{\rm num}}$({\color{black} \bf ---}) $\tilde{f}_{2}$({\color{KMgreen} \bf - -}) $\tilde{f}_{1}$({\color{KMpink} \bf - -}) } & 
\quad  \mbox{\footnotesize \bf (c) $f_{{\rm num}}$({\color{black} \bf ---}) $\tilde{f}_{2}$({\color{KMgreen} \bf - -}) $\tilde{f}_{1}$({\color{KMpink} \bf - -}) }         \\\\
\includegraphics[height=1.8in,width=2.3in]{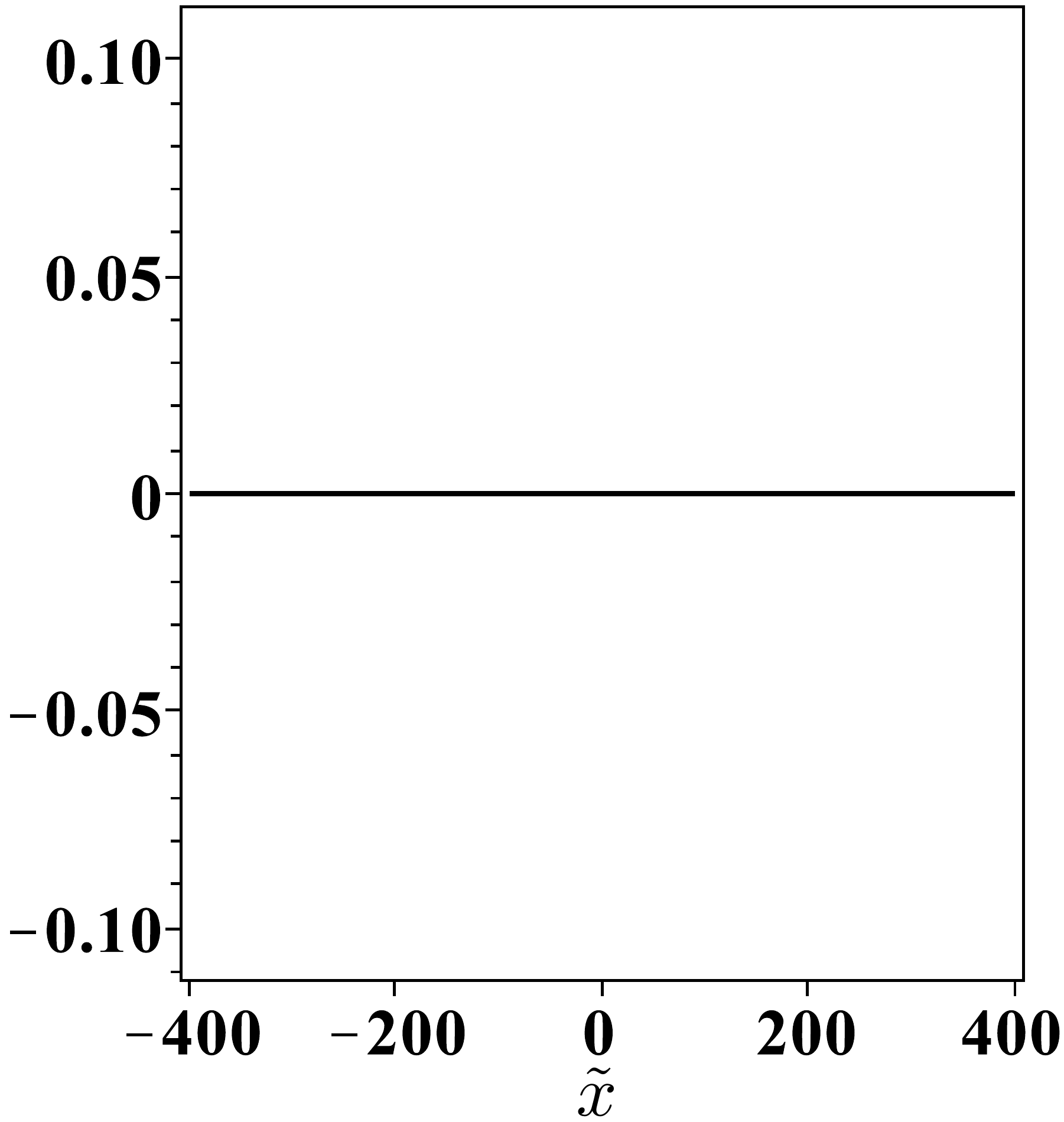}&
\includegraphics[height=1.8in,width=2.45in]{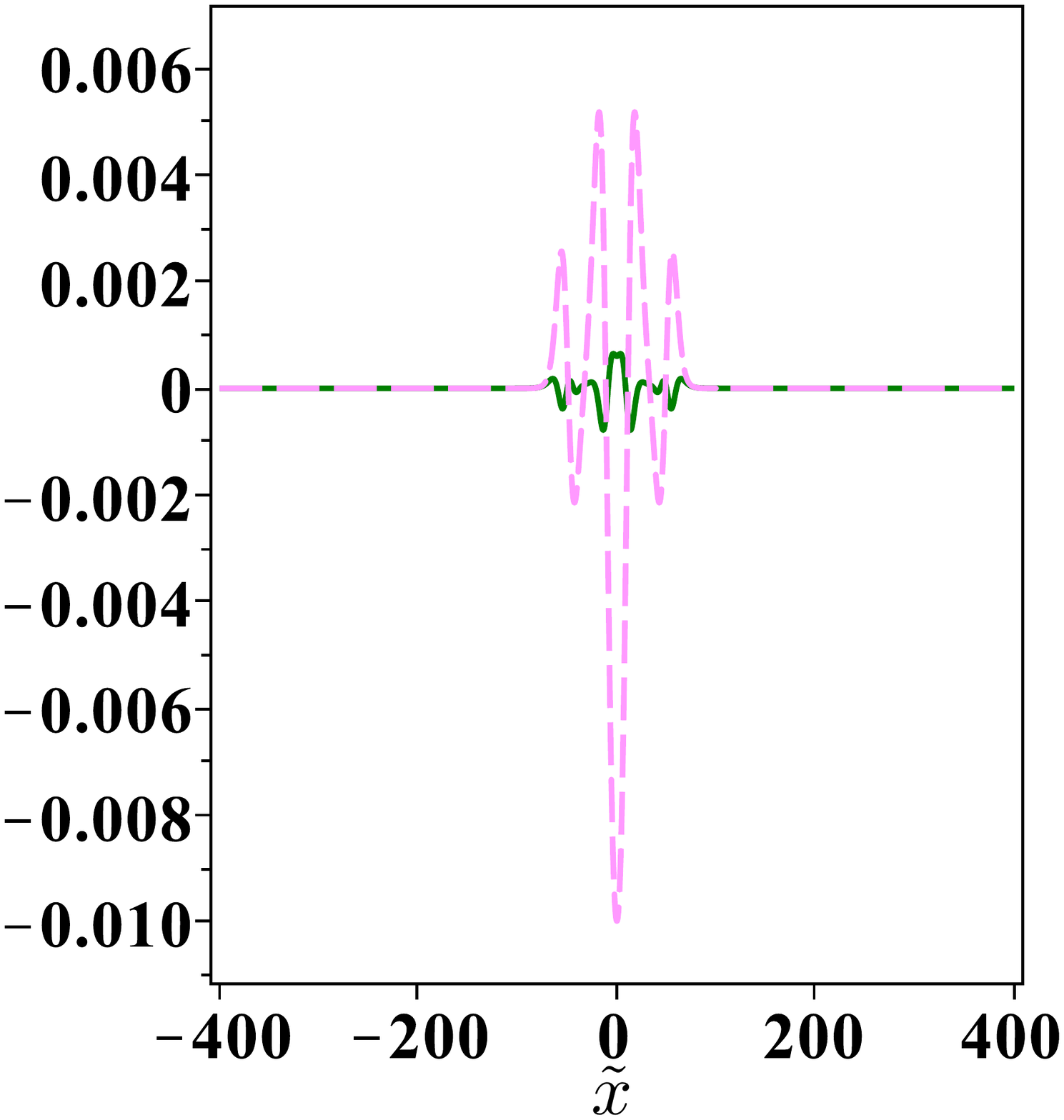} &                  
\includegraphics[height=1.8in,width=2.45in]{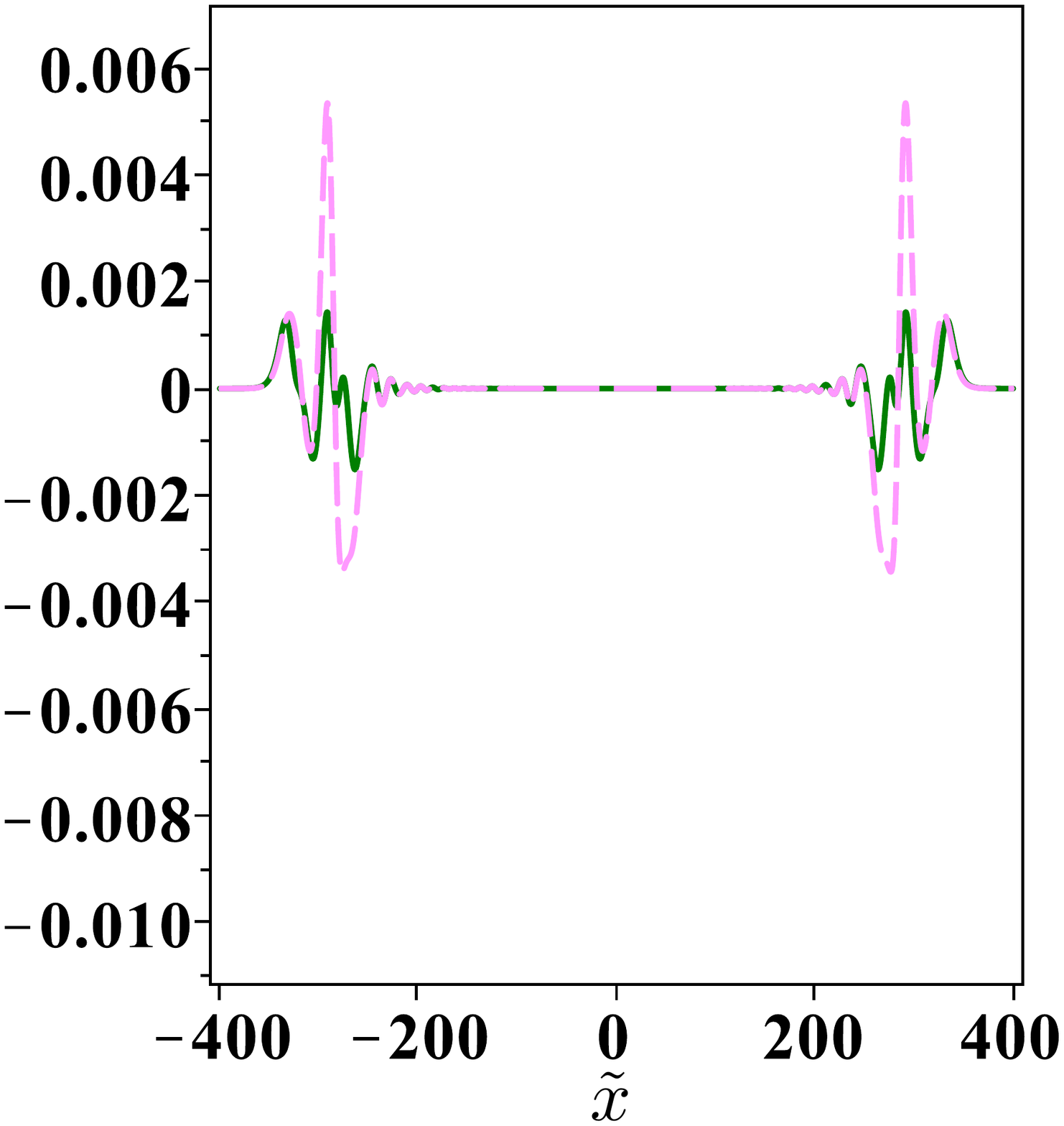}  \\              
\quad  \mbox{\footnotesize \bf (d) $\frac{\partial}{\partial t} (f_{{\rm num}}, \tilde{f}_{2},   \tilde{f}_{1})$({\color{black} \bf ---})  } &
\mbox{ \footnotesize \bf (e) \scriptsize  ($f_{{\rm num}}$  -  $\tilde{f}_{2}$)({\color{KMgreen} \bf ---})     \       ($f_{{\rm num}}$  -  $\tilde{f}_{1}$)({\color{KMpink} \bf - -}) }  & 
 \mbox{ \footnotesize \bf (f) \scriptsize  ($f_{{\rm num}}$  -  $\tilde{f}_{2}$)({\color{KMgreen} \bf ---})     \       ($f_{{\rm num}}$  -  $\tilde{f}_{1}$)({\color{KMpink} \bf - -}) }    \\
\end{array}$
\end{center}
\caption{\small Right- and left-propagating 2-soliton initial conditions (a) $\&$ (d). Evolution of the numerical solution compared with the weakly nonlinear solution $\tilde{f_2}$ and $\tilde{f_1}$ at (b) $\tilde{t}=32\ (\approx t_a)$, (c) $\tilde{t}=300\ (\approx t_b)$, and the absolute errors at each of the respective times (e) $\&$ (f). Other parameters are $\epsilon=0.1, k_1=0.61, k_2=0.56$ and $\hat{\alpha}_1=\hat{\alpha}_2=0$.} 
\label{figure:2sol_2way}
\end{figure}

For the example presented here we choose the same $\hat{\alpha}_i$'s and $k_i$'s as chosen in the previous example. The initial conditions of the IVP (\ref{eqn:IVP}) for this example are shown in Fig. \ref{figure:2sol_2way}a and  \ref{figure:2sol_2way}d, and the evolution of the weakly nonlinear solution, taken up to first and second order, along with the numerical solution, is shown in Fig. \ref{figure:2sol_2way}b and  \ref{figure:2sol_2way}c.
As the solution evolves one can notice qualitatively the resemblance of the leading order terms in the previous example propagating in both directions.
The absolute errors of the weakly nonlinear solution up to each order are shown in Fig. \ref{figure:2sol_2way}e and  \ref{figure:2sol_2way}f at each of the respective times and it's again clear that they are within the derived expected accuracy throughout the time interval considered.
There is a significant improvement in the accuracy of $\tilde{f}_2$ compared with $\tilde{f}_1$ for this example especially at earlier time, shown in Fig.  \ref{figure:2sol_2way}e, where $e^2_t/e^1_t$ is approximately $O(\epsilon)$.

\section{Perturbations of exactly solvable initial conditions}   \label{sec:1way_nsol_pert}

Finally we consider the weakly nonlinear solution for perturbations to the exactly solvable initial conditions of the IVP (\ref{eqn:IVP}) considered in Section \ref{sec:examples}. More specifically we examine just one particular case of initial conditions for a perturbation of a right propagating KdV N-soliton solution. The initial conditions of the IVP (\ref{eqn:IVP}) are chosen in the simplest perturbed form
\begin{eqnarray*}
 f|_{t=0} = 12 \frac{\partial^2}{\partial x^2} {\rm log [det} M_N(x,0)]  + \epsilon F^1(x) ,    \ \ \  \ \ \
 f_t|_{t=0} =-12 \frac{\partial^3}{\partial x^3} {\rm log [det} M_N(x,0)].
 \label{one_way_N_sol_pert_ICs}
\end{eqnarray*}
Therefore the leading order terms are still given in the form considered in Section \ref{sec:1way_nsol}
\begin{eqnarray*}
f^+=0, \qquad f^-= f^-_N(\xi,T),
\end{eqnarray*}
but the higher order terms are of the form
\begin{eqnarray*}
f^1 =  \frac 1 2 \left[R_1(\xi,T) + R_1 (\eta,T) -  \int^{\eta}_{\xi} R_2(x,T)dx \right], \quad \mbox{where}
\end{eqnarray*}
\begin{eqnarray*}
R_1(x,T)  = F^1(x),  \qquad    R_2(x,T) = \frac{\partial}{\partial T}f_N^-(x,T).
\end{eqnarray*}
Therefore the weakly nonlinear solution for a perturbation of the KdV N-soliton initial conditions propagating to the right, is as follows
\begin{eqnarray}
f=  12 \frac{\partial^2}{\partial \xi^2} {\rm log [ det} M_N(\xi,-T)]
+ \frac{\epsilon}{2} \left\lbrace
 F^1(\xi) + F^1(\eta)
-12 \left[ \frac{\partial^2}{\partial x \partial T} {\rm log [ det} M_N(x,-T)]  \right]^{x=\eta}_{x=\xi}  \right \rbrace   +  O(\epsilon^2).
\label{eqn:WNS_1way_pert}
\end{eqnarray}


For example, we consider a perturbation of the single KdV soliton solution as in Section \ref{sec:1way_1sol} and choose the perturbation to be defined by
\begin{eqnarray*}
F^1(x) = {\rm sech}\left(\frac{k x + \alpha}{2}    \right).
\end{eqnarray*}  
Therefore the initial conditions of the IVP are 
\begin{eqnarray}
f|_{t=0}= 12 \frac{\partial^2}{\partial x^2} {\rm log} (1+e^{\theta(x,0)})
+ \epsilon \ {\rm sech}\left(\frac{k x + \alpha}{2}    \right), \qquad 
f_t|_{t=0} = -12 \frac{\partial^3}{\partial x^3} {\rm log} (1+e^{\theta(x,0)})  ,
\label{eqn:1sol_1way_ICs_pert}
\end{eqnarray}
and from (\ref{eqn:WNS_1way_pert}) the weakly nonlinear solution is explicitly given by
\begin{eqnarray}
f &=& 3k^2{\rm sech}^2\left(\frac{k}{2}\left(\xi - \frac{k^2}{2}T\right) + \frac{\alpha}{2}   \right)
+ \frac{\epsilon}{2} \left\lbrace {\rm sech}\left(\frac{k\xi + \alpha}{2}\right) + {\rm sech}\left(\frac{k\eta + \alpha}{2}\right)     \right.
\nonumber \\
&+& \left. \frac{3k^4}{2}\left [-{\rm sech}^2  \left( \frac{k}{2}\left(\xi - \frac{k^2}{2}T\right)  + \frac{\alpha}{2}  \right)
+{\rm sech}^2\left(\frac{k}{2}\left(\eta - \frac{k^2}{2}T\right) + \frac{\alpha}{2} \right) \right]  \right \rbrace + O(\epsilon^2).
\label{eqn:1sol_1way_pert_WNS}
\end{eqnarray}
To analyse the error we transform the variables to those used in the numerics and for simulations we use the following initial conditions in order to comply with (\ref{eqn:1sol_1way_ICs_pert})
\begin{eqnarray*}
f_{i,0} &=& 3k^2 \epsilon\  \text{sech} ^2 \left(\frac{k\sqrt{\epsilon} {\tilde{x}} + \alpha}{2}\right) 
+  \epsilon^2  \text{sech} \left(\frac{k\sqrt{\epsilon} {\tilde{x}} + \alpha}{2}\right) , 
\nonumber \\
f_{i,1} &=&  3k^2\epsilon\  \text{sech} ^2  \left(\frac{k\sqrt{\epsilon} ({\tilde{x}}-\kappa) + \alpha}{2}\right) 
+  \frac{\epsilon^2}{2} \left \lbrace \text{sech} \left(\frac{k\sqrt{\epsilon} ({\tilde{x}}-\kappa) + \alpha}{2}\right)     
+   \text{sech} \left(\frac{k\sqrt{\epsilon} ({\tilde{x}}+ \kappa) + \alpha}{2}\right)
 \right \rbrace.
\end{eqnarray*}

\begin{figure}[h]
\begin{center}$
\begin{array}{ccc}
\quad \tilde{t}=0 & \quad \tilde{t}=32&  \quad \tilde{t}=300 \\
&&\\
\includegraphics[width=2.1in]{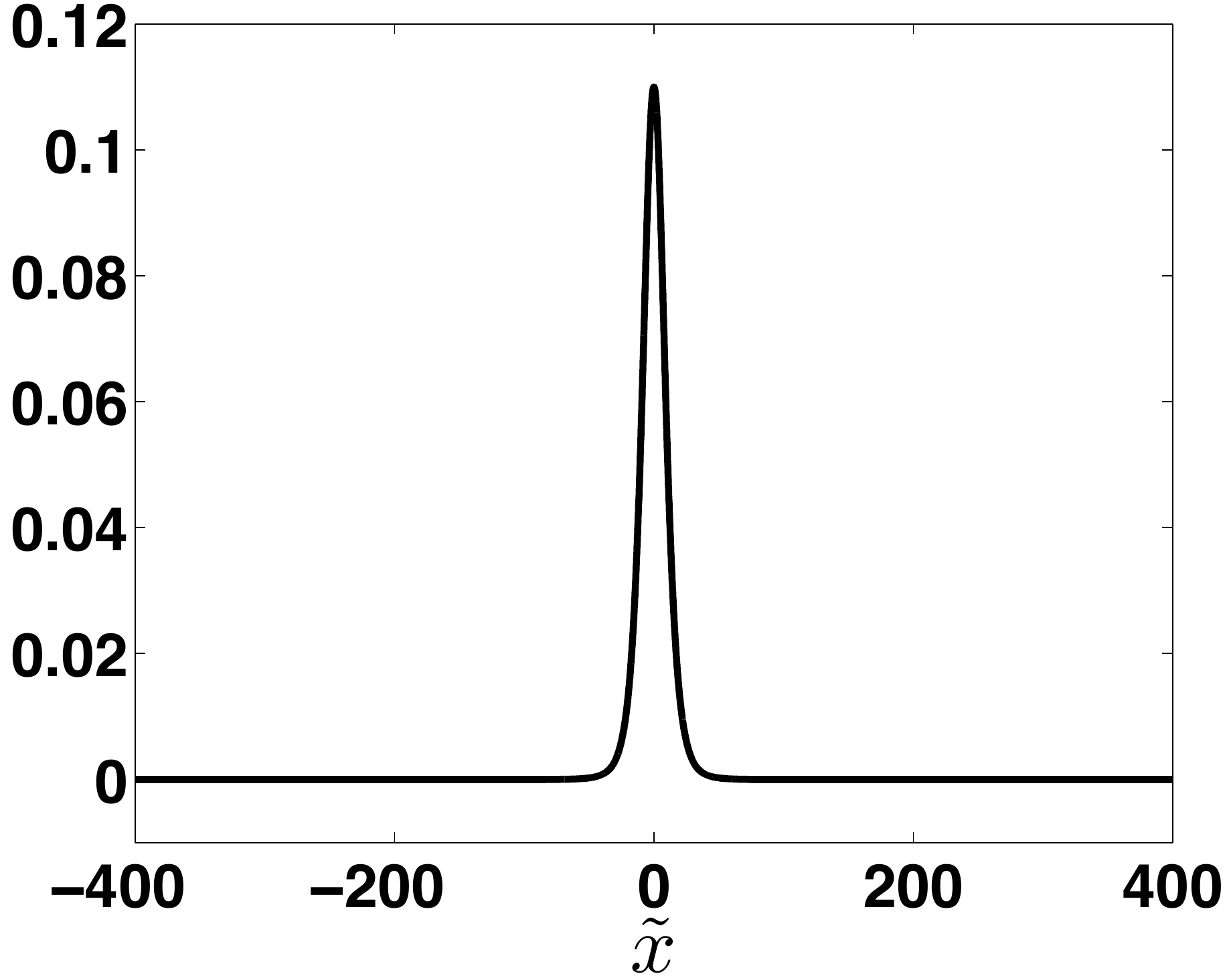} &
\includegraphics[width=2.1in]{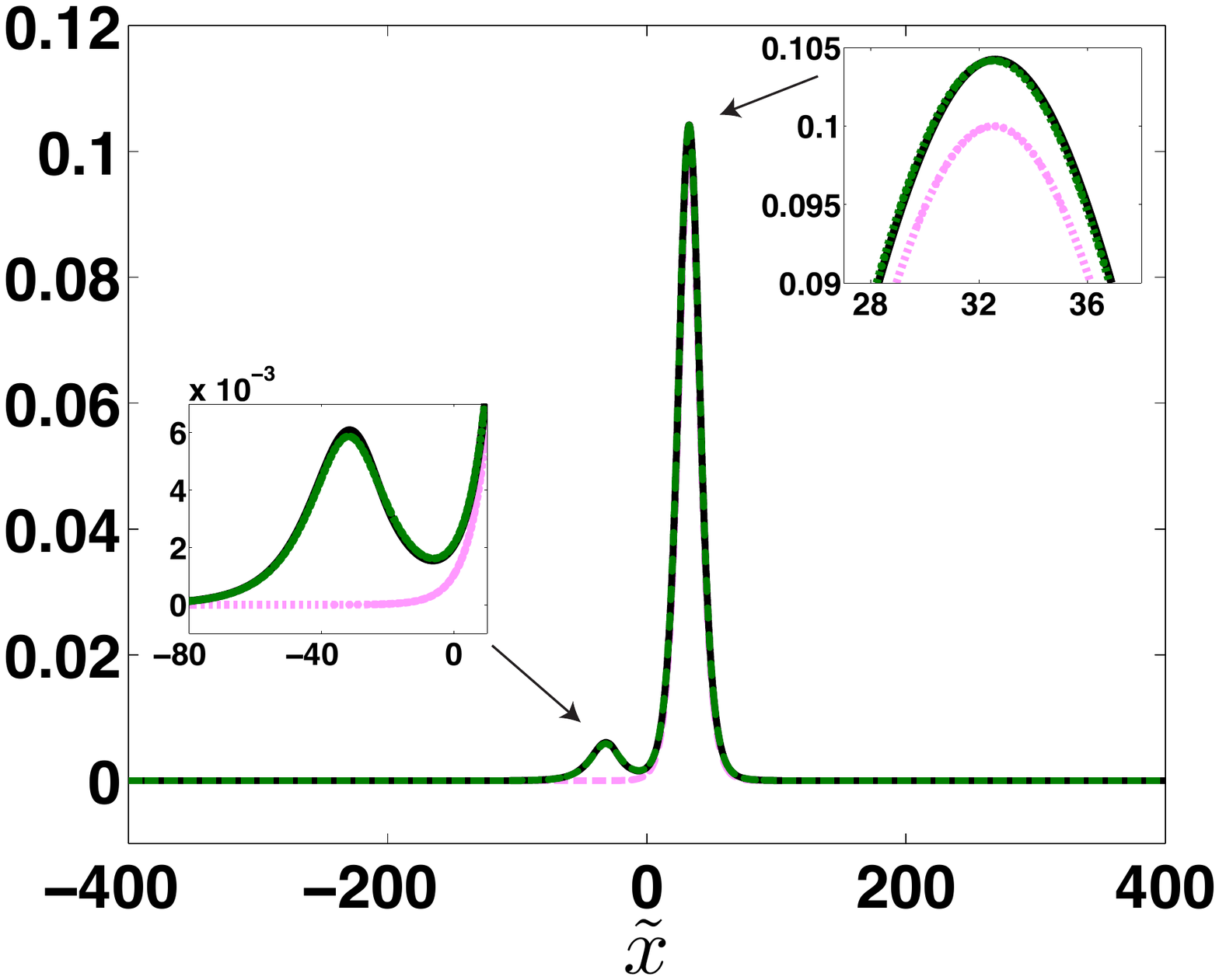} &    
\includegraphics[width=2.1in]{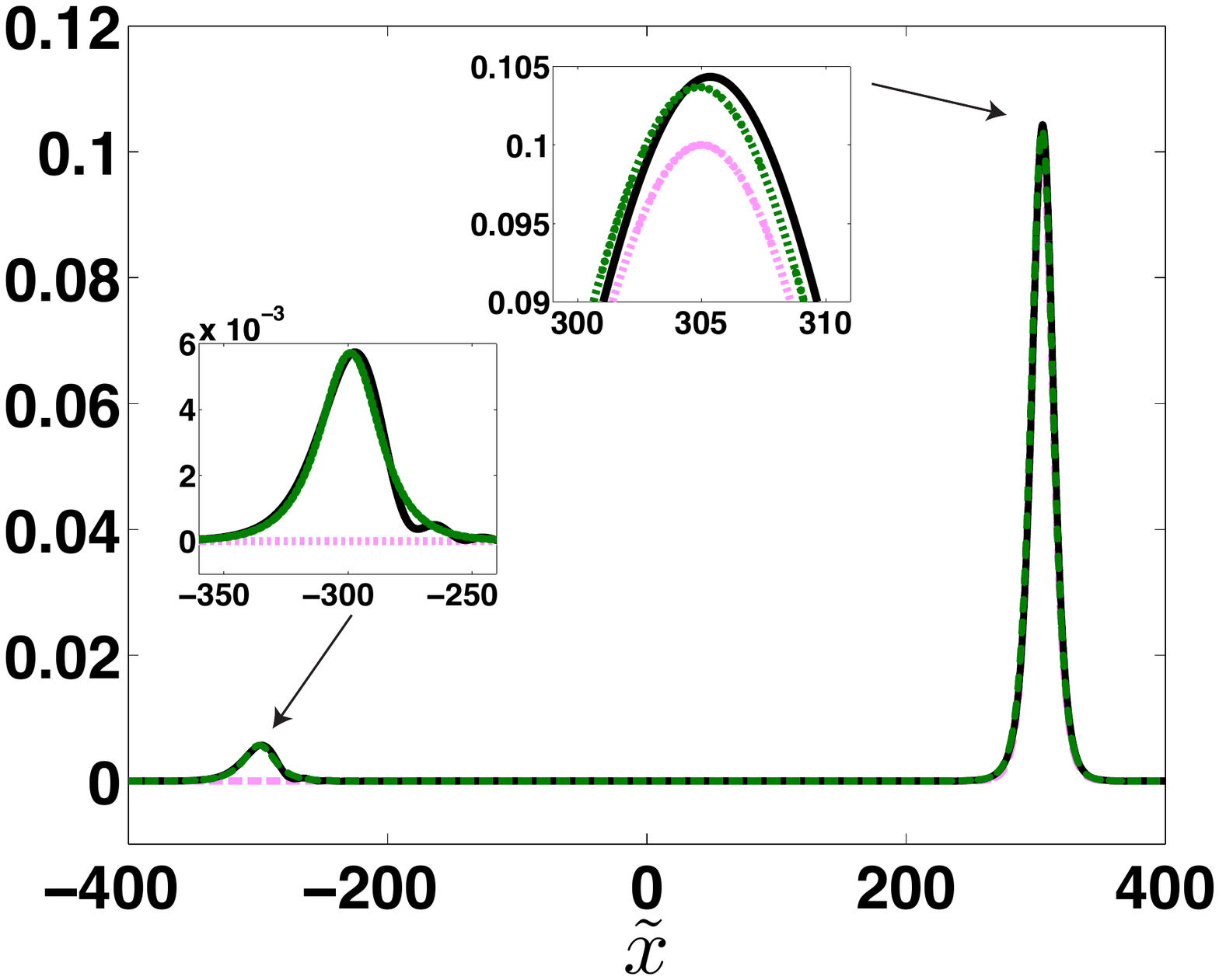} \\             
\quad  \mbox{\footnotesize \bf (a) $f_{{\rm num}}, \tilde{f}_{2},   \tilde{f}_{1}$({\color{black} \bf ---})  }  &
\quad  \mbox{\footnotesize \bf (b) $f_{{\rm num}}$({\color{black} \bf ---}) $\tilde{f}_{2}$({\color{KMgreen} \bf - -}) $\tilde{f}_{1}$({\color{KMpink} \bf - -}) } & 
\quad  \mbox{\footnotesize \bf (c) $f_{{\rm num}}$({\color{black} \bf ---}) $\tilde{f}_{2}$({\color{KMgreen} \bf - -}) $\tilde{f}_{1}$({\color{KMpink} \bf - -}) }         \\\\
\includegraphics[height=1.7in,width=2.3in]{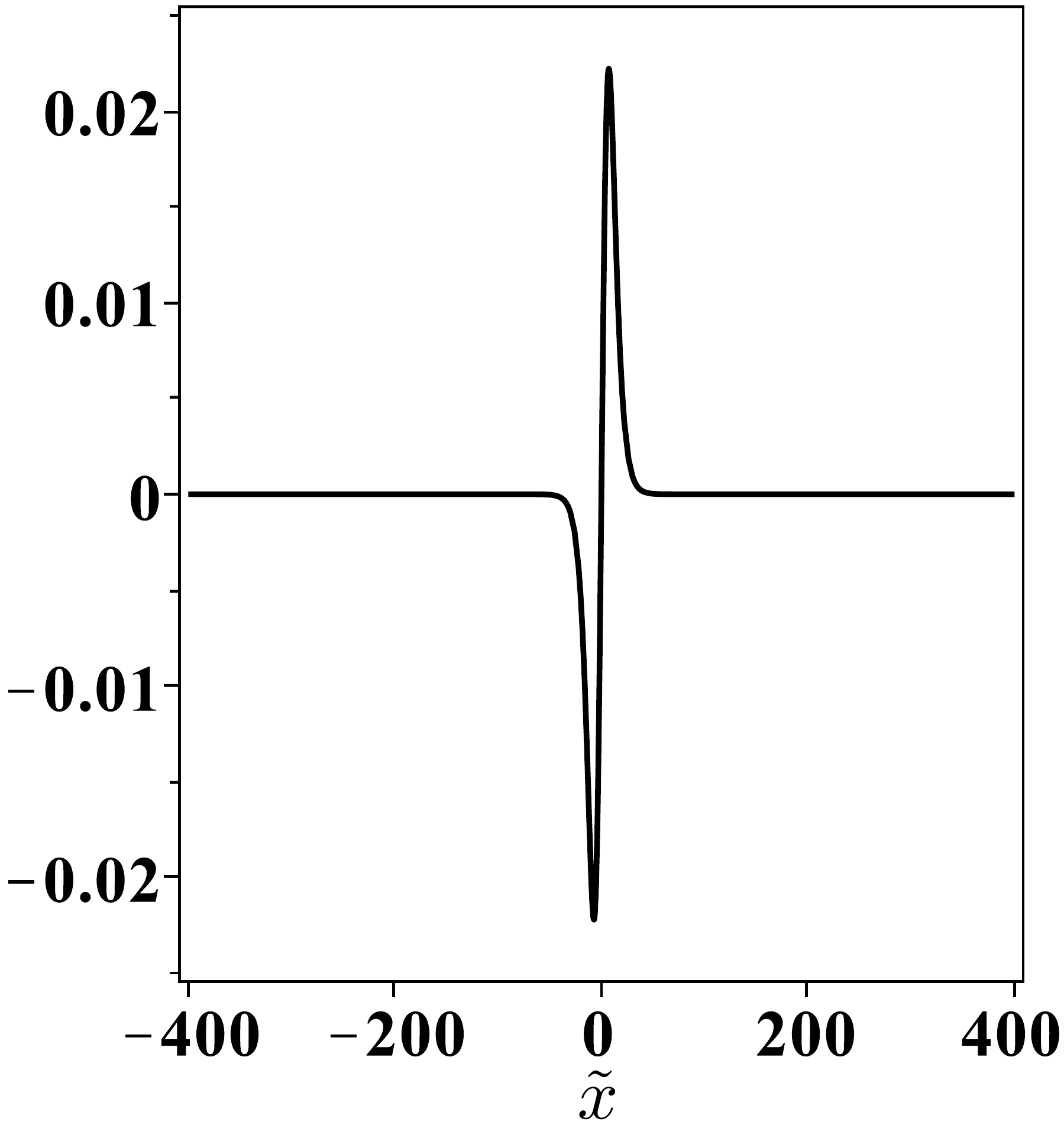} &    		
\includegraphics[width=2.1in]{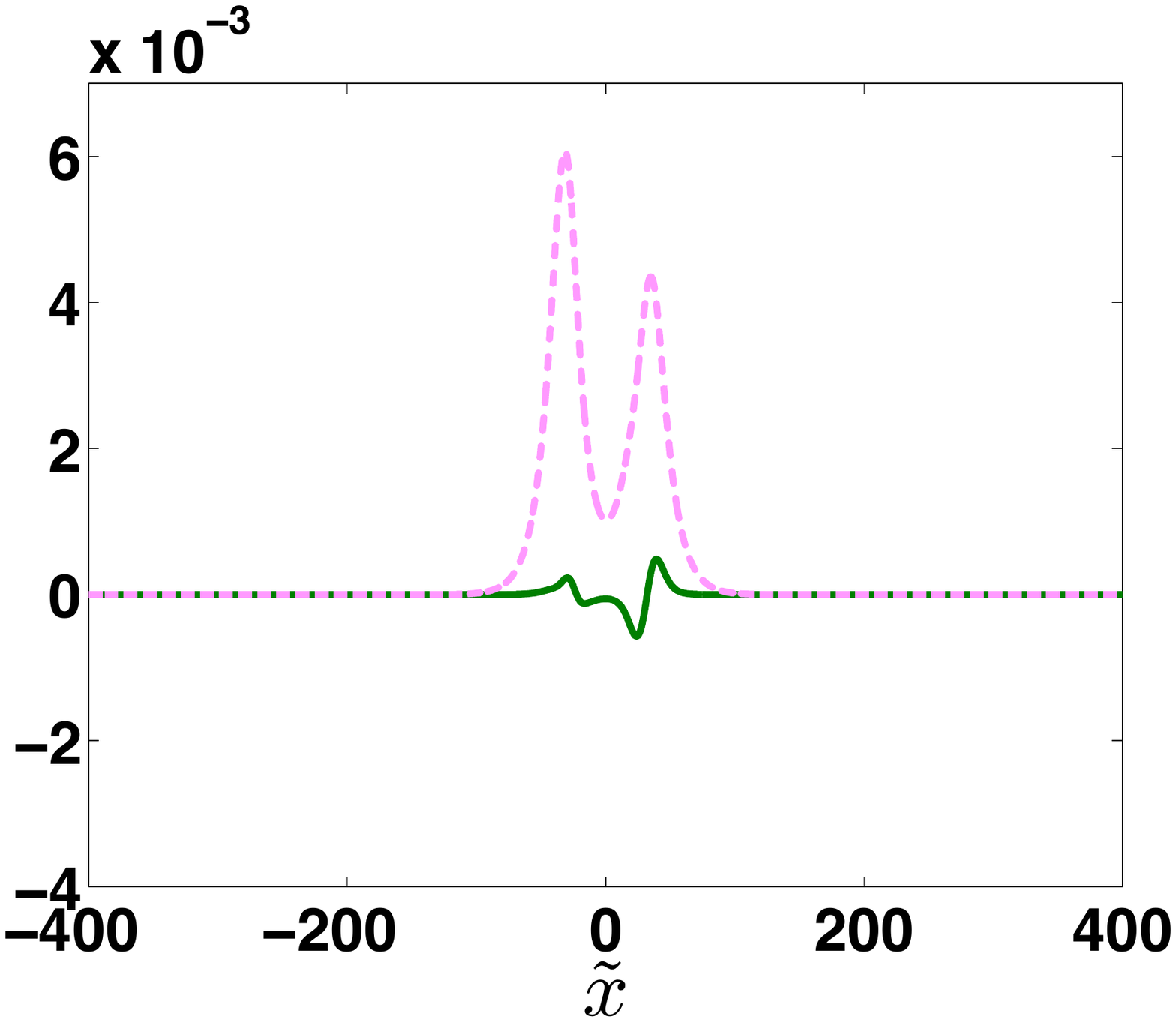} & 		
\includegraphics[width=2.1in]{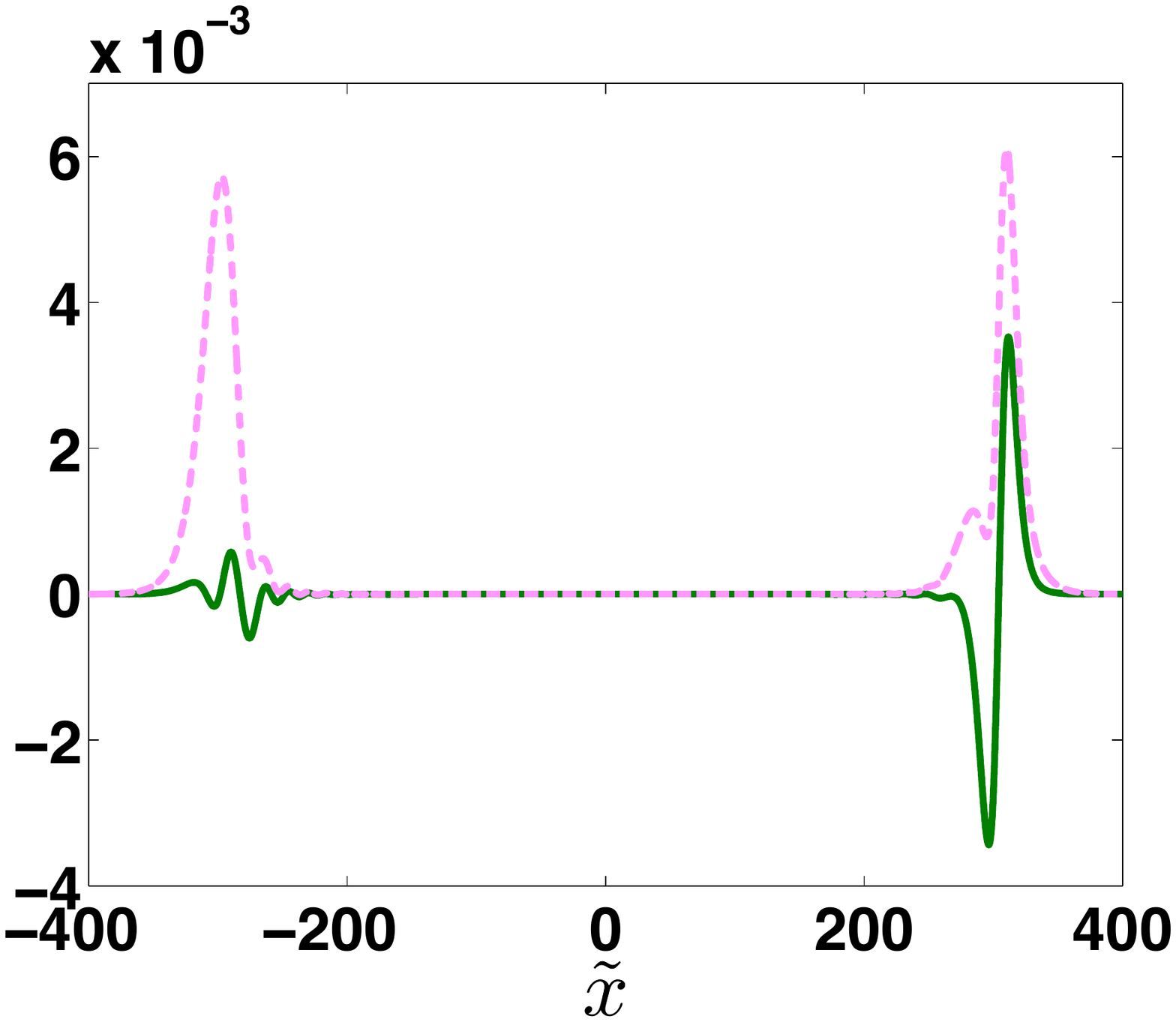} \\		
\quad  \mbox{\footnotesize \bf (d) $\frac{\partial}{\partial t} (f_{{\rm num}}, \tilde{f}_{2},   \tilde{f}_{1})$({\color{black} \bf ---})  } &
\mbox{ \footnotesize \bf (e) \scriptsize  ($f_{{\rm num}}$  -  $\tilde{f}_{2}$)({\color{KMgreen} \bf ---})     \       ($f_{{\rm num}}$  -  $\tilde{f}_{1}$)({\color{KMpink} \bf - -}) }  & 
 \mbox{ \footnotesize \bf (f) \scriptsize  ($f_{{\rm num}}$  -  $\tilde{f}_{2}$)({\color{KMgreen} \bf ---})     \       ($f_{{\rm num}}$  -  $\tilde{f}_{1}$)({\color{KMpink} \bf - -}) }    \\
\end{array}$
\end{center}
\caption{\small Perturbation of right-propagating 1-soliton initial conditions (a) $\&$ (d). Evolution of the numerical solution compared with the weakly nonlinear solution $\tilde{f_2}$ and $\tilde{f_1}$ at (b) $\tilde{t}=32\ (\approx t_a)$, (c) $\tilde{t}=300\ (\approx t_b)$, and the absolute errors at each of the respective times (e)$\ \&$ (f). All other parameters are $\epsilon=0.1,\ k=\frac{1}{\sqrt{3}}$ and $\alpha=0$.}  
\label{figure:1way_1sol_pert}
\end{figure}

The initial conditions of the IVP (\ref{eqn:IVP}), for this example, are shown in Fig. \ref{figure:1way_1sol_pert}a and \ref{figure:1way_1sol_pert}d, and the evolution of the weakly nonlinear solution, up to leading and second order, along with the numerical solution are shown in Fig. \ref{figure:1way_1sol_pert}b and \ref{figure:1way_1sol_pert}c. Figures \ref{figure:1way_1sol_pert}e and \ref{figure:1way_1sol_pert}f depict the corresponding errors of the weakly nonlinear solution at each of the respective times. The weakly nonlinear solution (\ref{eqn:1sol_1way_pert_WNS}) is within its derived accuracy throughout the time interval considered and is significantly more accurate than the leading order solution, most evident for early time (Fig. \ref{figure:1way_1sol_pert}e). However as time increases beyond the region of validity of the weakly nonlinear solution we find that $e^2_{t} \rightarrow e^1_{t}$.

\section{Conclusion}

In this paper we have constructed a nonsecular weakly nonlinear solution of the IVP for the Boussinesq equation on the infinite line, for initial data generating sufficiently rapidly decaying right- and left-propagating waves. Seeking asymptotic multiple-scale expansions and implementing an averaging procedure with respect to fast time we derived two KdV equations describing the leading order terms and obtained formulae for the higher order corrections in an explicit and simple form.
The initial data was split into $O(1)$ and $O(\epsilon)$ parts, and it was shown, in a case study, that this allows one to obtain explicit approximate solutions of the IVP for `exactly solvable initial conditions' (from the viewpoint of the Inverse Scattering Transform for the leading order KdV equations).

A finite difference scheme was implemented primarily to measure the accuracy of the developed weakly nonlinear solution.
We showed through comparisons of the numerical solution with an exact solution of the Boussinesq equation that the scheme's accuracy was well within the required accuracy to measure the weakly nonlinear solution. Albeit that changing the initial conditions for simulations would undoubtedly alter the magnitude of the error of the numerical scheme determined in the test case in Section \ref{sec:numerics}, it remains a good indicator of the scheme's accuracy.

%
%
On comparison of the weakly nonlinear solution with relevant numerical simulations there arose two consistent features in each of the examples considered. Firstly, the weakly nonlinear solution remained within its required accuracy throughout the time interval of its validity. For a more accurate solution and applicability for larger time one must reformulate equation (\ref{eqn:bous}) to consider higher order terms.
Secondly, the maximum absolute error for our weakly nonlinear solution ($e^2_t$) was significantly lower than for the leading order solution ($e^1_t$). This was particularly evident for earlier times. 
As time extended beyond the end of the validity of the weakly nonlinear solution it can be seen that $e^2_t$ and $e^1_t$ become comparable. 

To conclude, in all examples considered in this paper the constructed weakly nonlinear solution was in excellent agreement with the results of direct numerical simulations within the range of its asymptotic validity. It would be interesting to consider other classes of initial conditions for the leading order KdV equations, using the well-developed techniques of the Inverse Scattering Transform. 
It is also desirable to implement a similar procedure for water waves, which can be done along the same lines. Finally, one can explore the possibilities of renormalising the constructed solution, similar to \cite{Gear}, which will be reported elsewhere.

\section{Acknowledgments}

We thank C. Klein and A.S. Topolnikov for helpful advice on numerical simulations, and R.H.J. Grimshaw for pointing at several useful ideas in \cite{Gear} and for helpful discussions.


\begin{thebibliography}{99}

\bibitem{AS} ABLOWITZ, M. J. \& SEGUR, H. (1981) {\it Solitons and the Inverse Scattering Transform}. SIAM Philadelphia.

\bibitem{thomas} AMES, W. F. (1979)  {\it Numerical Methods for Partial Differential Equations}. Academic Press, Inc, Thomas Nelson and Sons Ltd.

\bibitem{BYC} BEN YOUSSEF, W.  \& COLIN, T. (2000) Rigorous derivation of Korteweg-de Vries-type systems from a general class of nonlinear hyperbolic systems. {\it M2AN Math. Model. Numer. Anal.}, {\bf34}, 873-911.


\bibitem{BBM} BENJAMIN, T. B., BONA, J. L. \& MAHONY, J. J. (1972) Model equations for long waves in nonlinear dispersive systems. {\it Philos. Trans. R. Soc. Lond. A}, {\bf272}, 47-48.


\bibitem{BONA1} BONA, J. L., CHEN, M. \& SAUT, J.-C. (2002) Boussinesq equations and other systems for small-amplitude long waves in nonlinear dispersive media. I: Derivation and linear theory. {\it J. Nonlinear Sci.}, {\bf12}, 283-318.

\bibitem{BONA2} BONA, J. L., CHEN, M. \& SAUT, J.-C. (2004) Boussinesq equations and other systems for small-amplitude long waves in nonlinear dispersive media. II: The nonlinear theory. {\it Nonlinearity}, {\bf 17}, 925-952.


\bibitem{BCL} BONA, J. L., COLIN, T. \& LANNES, D. (2005) Long wave approximations for water waves. {\it Arch. Rational Mech. Anal.}, {\bf178}, 373-410.

\bibitem{BOUS}  BOUSSINESQ, J. V. (1872) Th\'{e}orie des ondes et des remous qui se propagent le long d'un canal rectangulaire horizontal, en communiquant au liquide contenu dans ce canal des vitesses sensiblement pareilles de la surface au fond. {\it J. Math. Pures Appl., ser. (2)}, {\bf17}, 55-108.

\bibitem{P9} BRATOS, A. G. (2009) A predictor-corrector scheme for the improved Boussinesq equation. {\it Chaos, Solitons and Fractals}, {\bf40}, 2083-2094.

\bibitem{Christov} CHRISTOV, C. I.,  MAUGIN, G. A. \& VELARDE, M. G. (1996) Well-posed Boussinesq paradigm with purely spatial higher-order derivatives. {\it Phys. Rev. E}, {\bf54}, 3621-3638.

\bibitem{Craig} CRAIG, W. (1985)  An existence theory for water waves and the Boussinesq and Korteweg-de Vries scaling limits. {\it Comm. Part. Diff. Eqs}, {\bf10}, 787-1003.

\bibitem{DJ} DRAZIN, P. G. \& JOHNSON, R. S. (1989) {\it Solitons: an introduction}. Cambridge University Press, Cambridge.

\bibitem{P4} EL-ZOHEIRY, H. (2003) Numerical investigation for the solitary waves interaction of the ``good" Boussinesq equation. {\it Appl. Numer. Math.}, {\bf45}, 161-173.

\bibitem{FPU} FERMI, E., PASTA, J. \& ULAM, S. (1955) Studies on nonlinear problems, I. {\it Los Alamos Scientific Laboratory Report No. LA-1940}. 
\\
NEWELL, A.C. (ed.) (1974) Nonlinear Wave Motion. {\it AMS Lect. Appl. Math.}, {\bf15}, 143-156.

\bibitem{FP} FOKAS, A. S. \& PELLONI, B. (2005) Boundary value problems for Boussinesq type systems. {\it Math. Phys. Anal. Geom.}, {\bf8}, 59-96.

\bibitem{GGKM} GARDNER, C. S., GREENE, J. M., KRUSKAL, M. D.  \& MIURA, R. M. (1967) Method for solving the Korteweg-de Vries equation. {\it Phys. Rev. Lett.}, {\bf19}, 1095-1097.

\bibitem{Gear} GEAR, J. \& GRIMSHAW, R. (1984) Weak and strong interactions between internal solitary waves. {\it Stud. Appl. Math.}, {\bf70}, 235-258.

\bibitem{P6} HAJJI, M. A. \& AL-KHALED, K. (2007) Analytic studies and numerical simulations of the generalized Boussinesq equation. {\it Appl. Math. and Comp.}, {\bf191}, 320-333.

\bibitem{Hirota} HIROTA, R. (1971) Exact solution of the Korteweg-de Vries equation for multiple collisions of solitons. {\it Phys. Rev. Lett.}, {\bf27}, 1192-1194. 

\bibitem{P11} IRK, D. \& DAG, I. (2009) Numerical simulations of the improved Boussinesq equation. {\it Numer. Meth. for Partial Diff. Equations}, {\bf26}, 1316-1327.

\bibitem{JE} JANNO, J. \& ENGELBRECHT, J. (2005) Solitary waves in nonlinear microstructured materials. {\it J. Phys. A: Math. Gen.}, {\bf38}, 5159-5172.

\bibitem{J} JOHNSON, R. S. (1997) {\it A Modern Introduction to the Mathematical Theory of Water Waves}. Cambridge University Press, Cambridge.



\bibitem{KL} KALANTAROV, V.K. \& LADYZHENSKAYA, O. A. (1978) The occurrence of collapse for quasilinear equations of parabolic and hyperbolic types. {\it J. Sov. Math.}, {\bf10}, 53-70.

\bibitem{K} KALYAKIN, L. A. (1989) Long-wave asymptotics. Integrable equations as the asymptotic limit of nonlinear systems. {\it Uspekhi Mat. Nauk}, {\bf44}, 5-34.

\bibitem{KN} KANO, T. \& NISHIDA, T. (1986) A mathematical justification fro Korteved-de Vries equation and Boussinesq equation of water surface waves. {\it Osaka J. Math.}, {\bf23}, 389-413.

\bibitem{KayM} KAY, I. \& MOSES, H. E. (1956) Reflectionless transmission through dielectrics and scattering potentials. {\it J. Appl. Phys.}, {\bf27}, 1503-1508.

\bibitem{KM} KHUSNUTDINOVA, K. R. \& MOORE, K. R. (2011) Initial-value problem for coupled Boussinesq equations and a hierarchy of Ostrovsky equations. {\it Wave Motion}, {\bf48}, 738-752.

\bibitem{Fission} KHUSNUTDINOVA, K. R. \& SAMSONOV, A. M. (2008) Fission of a longitudinal strain solitary wave in a delaminated bar. {\it Phys. Rev. E}, {\bf77}, 066603.

\bibitem{KSZ} KHUSNUTDINOVA, K. R., SAMSONOV, A. M. \& ZAKHAROV, A. S. (2009) Nonlinear layered lattice model and generalized solitary waves in layered elastic structures. {\it Phys. Rev. E}, {\bf79}, 056606.

\bibitem{KDV} KORTEWEG, D. J. \& DE VRIES, G. (1895) On the change of form of long waves advancing in a rectangular canal, and on a new type of long stationary waves. {\it Philos. Mag. (5)}, {\bf39}, 422-443.

\bibitem{Lannes} LANNES, D. (2005) Well-Posedness of the Water Waves Equations. {\it J. Am. Math. Soc.}, {\bf18}, 605-654. 

\bibitem{Maugin} MAUGIN, G. A. (1999) {\it Nonlinear Waves in Elastic Crystals}. Oxford University Press, Oxford.

\bibitem{Miles1} MILES, J.W. (1977) Obliquely interacting solitary waves. {\it J. Fluid Mech.}, {\bf79}, 157-169.

\bibitem{P3} MOHSEN, A., EL-ZOHEIRY, H. \& ISKANDAR, L. (1993) A highly accurate finite-difference scheme for a Boussinesq-type equation. {\it Appl. Math. and Comp.}, {\bf55}, 201-212.

\bibitem{Ostrovsky} OSTROVSKY, L. A. (1978) Nonlinear internal waves in a rotating ocean. {\it Oceanology}, {\bf18}, 119-125.

\bibitem{Porubov} PORUBOV, A. V. (2003) {\it Amplification of Nonlinear Strain Waves in Solids}. World Scientific, Singapore.


\bibitem{Samsonov}  SAMSONOV, A. M. (2001) {\it Strain Solitons in Solids and How to
Construct Them}. Chapman and Hall/CRC, Boca Raton.

\bibitem{S} SCHNEIDER, G. (1998) The long wave limit for a Boussinesq equation. {\it SIAM J. Appl. Math.}, {\bf58}, 1237-1245.

\bibitem{SW} SCHNEIDER, G. \& WAYNE, C. E. (2000) The long-wave limit for the water wave problem. I. The case of zero surface tension. {\it Comm. Pure Appl. Math.}, {\bf53}, 1475-1535.
%
%
%
\bibitem{SOR} SOERENSEN, M. P., CHRISTIANSEN, P. L. \& LOMDAHL, P. S. (1984) Solitary waves on nonlinear elastic rods.I. {\it J. Acoust. Soc. Am.}, {\bf76}, 871-879.


\bibitem{WW} WAYNE, C. E. \& WRIGHT, J. D. (2002) Higher order modulation equations for a Boussinesq equation. {\it SIAM J. Appl. Dyn. Syst.}, {\bf1}, 271-302.


\bibitem{ZK} ZABUSKY, N. J. \& KRUSKAL, M. D. (1965) Interaction of solitons in a collisionless plasma and the recurrence of initial states. {\it Phys. Rev. Lett.}, {\bf15}, 240-243.




\bibitem{ZAK} ZAKHAROV, V. E. (1974) On stochastisation of one-dimensional chains of nonlinear oscillators. 
{\it Sov. Phys. JETP}, {\bf38}, 108-110.




































































































































\end{thebibliography}
\end{document}